\documentclass[10pt]{article} 
\pdfoutput=1  
\usepackage[top=1in, bottom=1in, left=1in, right=1in]{geometry}
\usepackage{times} 
\usepackage{xspace}
\usepackage{url}
\usepackage{amsmath, amssymb, esint}
\usepackage{xcolor}
\usepackage{graphicx}
\usepackage{subfig}
\usepackage{mathabx}
\usepackage[export]{adjustbox}

\usepackage[format=plain,justification=centering,singlelinecheck=false,font=small,labelfont=bf,labelsep=space]{caption}
\usepackage{bm}

\usepackage[hyperfootnotes=false]{hyperref}
\hypersetup{
    colorlinks=true,
    citecolor=blue,
    linkcolor=blue,
    filecolor=magenta,      
    urlcolor=blue
    }
\usepackage{cleveref}

\usepackage[utf8]{inputenc}
\usepackage[backend=biber,style=nature]{biblatex}
\usepackage{setspace}

\newcommand{\rb}{{\rm b}}
\newcommand{\rw}{{\rm w}}

\usepackage{enumitem}
\usepackage{wrapfig}
\usepackage{authblk}

\usepackage[outdir=./]{epstopdf}
\usepackage{longtable}
\usepackage{multicol}
\usepackage{ragged2e}
\DeclareRefcontext{supp}{labelprefix=S} 

\title{Fundamental Microscopic Properties as Predictors of Large-Scale Quantities of Interest: Validation through Grain Boundary Energy Trends}

\author[1]{Benjamin A. Jasperson}
\author[2]{Ilia Nikiforov}
\author[3]{Amit Samanta}
\author[4]{Brandon Runnels}
\author[1,5]{Harley T. Johnson}
\author[2]{Ellad B. Tadmor\footnote{corresponding author: tadmor@umn.edu}}

\affil[1]{Department of Mechanical Science and Engineering, University of Illinois Urbana-Champaign}
\affil[2]{Department of Aerospace Engineering and Mechanics, University of Minnesota}
\affil[3]{Lawrence Livermore National Laboratory}
\affil[4]{Department of Aerospace Engineering, Iowa State University}
\affil[5]{Materials Research Lab, University of Illinois Urbana-Champaign}

\date{\vspace{-1cm}}

\begin{document}

\maketitle

\begin{abstract}
Correlations between fundamental microscopic properties computable from first principles, which we term \textit{canonical properties}, and complex large-scale quantities of interest (QoIs) provide an avenue to predictive materials discovery. We propose that such correlations can be efficiently discovered through simulations utilizing approximate interatomic potentials (IPs), which serve as an ensemble of ``synthetic materials.'' 
As a proof of principle we build a regression model relating canonical properties to the symmetric tilt grain boundary (GB) energy curves in face-centered cubic crystals, characterized by the scaling factor in the universal lattice matching model of Runnels et al.\ (2016), which we take to be our QoI. Our analysis recovers known correlations of GB energy to other properties and discovers new ones.
We also demonstrate, using available density functional theory (DFT) GB energy data, that the regression model constructed from IP data is consistent with DFT results, confirming the assumption that the IPs and DFT belong to same statistical pool and thereby validating the approach.
Regression models constructed in this fashion can be used to predict large-scale QoIs based on first-principles data and provide a general method for training IPs for QoIs beyond the scope of first-principles calculations.
	
\end{abstract}

\begin{refsection}


\section{Introduction}
\label{sec:intro}

Materials science researchers have long used correlations between properties to elucidate the underlying physical mechanisms of materials. Examples include Vegard's rule \cite{Vegard1,Vegard2} (relating lattice parameter to composition), Hume-Rothery rules \cite{HumeRothery1,HumeRothery2,HumeRothery3,HumeRothery4} (describing the conditions under which an element is soluble in a metal), the Lindemann criterion \cite{Lindemann10} (relating melting temperature to phonon vibration amplitudes), the universal binding energy relation (UBER)~\cite{UBER, UBERorigin} (relating the volumetric dependence of the binding energy to the equilibrium atomic volume and bulk modulus), grain boundary (GB) energy rules \cite{udlerGrainBoundarySurface1996,suzukiInteractionPointDefects2003,suzukiDiffusionMechanismsGrain2004,holmComparingGrainBoundary2010,hallilCorrelationVacancyFormation2016}
(relating grain boundary energy to properties like the shear modulus and lattice constant), and so on. Such relations are of particular interest when they relate complex large-scale quantities of interest (QoIs) with fundamental microscopic properties, such as lattice parameters, elastic constants, surface energies, etc., that can be computed using accurate first-principles methods based on quantum mechanics.
We refer to these basic properties as ``canonical properties'' to reflect their fundamental nature.
Correlations between large-scale QoIs and canonical properties provide a pathway to predictive computational materials discovery.
In this paper, we propose that such relations can be inferred
in a statistical fashion from the information generated by a large number of atomistic simulations using classical interatomic potentials (IPs). The set of IPs used in these simulations constitutes an ensemble of ``virtual synthetic materials'' that together capture real material behavior.
We provide a proof of principle through atomistic simulations of symmetric tilt GB energies in face-centered cubic (FCC) metals where direct comparison with first-principles density functional theory (DFT) results is possible in some cases.

Classical IPs are fast,\footnote{Relative to first-principles methods.} approximate models of atomic interactions that are designed to reproduce the bonding patterns in targeted material classes.
In physics-based IPs, this information is encoded explicitly in the functional form.
For example, the Lennard-Jones pair potential is a good model for van der Waals interactions; the embedded atom method (EAM) potential provides a good description of metallic bonding; three-body, modified EAM (MEAM), and bond-order potentials capture covalent bonding; and so on \cite{tadmor:miller:2011}. 
In recent years, data-driven IPs using machine learning (ML) techniques have been growing in popularity \cite{Mishin2021}. Here universal approximators, such as Gaussian process regression, neural networks, and graph neural networks, are trained on large amounts of first-principles data (atomic configurations, energies, forces, etc.) that implicitly contain information about bonding. 
Neither approach is perfect. Physics-based IPs introduce a bias through their functional form. This can improve transferability (i.e., the ability to make reasonable predictions far from the IP's training data), however the restrictions imposed by the functional form limit IP accuracy. In contrast, ML-based IPs can accurately reproduce first-principles values for configurations close to those that an IP was trained on, but provide unreliable estimates far from the training data due to the lack of physics in the functional form. 

Although both IP classes are limited, we argue that by virtue of their functional form and/or training data, which are based on the domain expertise of the IP developer, many of these models encode the behavior of the classes of materials for which they are designed (when used within their domain of applicability).
Thus, even if an IP is not a perfect match for the material it is nominally modeling, say FCC nickel, it is a \textit{synthetic material} in the FCC class, obeying similar physics to those of real nickel.
The large number of IPs that have been developed over the years provide an ensemble of synthetic materials that can be used to discover correlations between material properties through statistical regression. This approach has been made viable through the advent of centralized repositories of IPs and their predictions for different material properties \cite{tadmor:elliott:2011, beckerConsiderationsChoosingUsing2013,Choudhary2020}. In particular, the OpenKIM framework \cite{tadmor:elliott:2011, openkim:website} includes a large collection of IPs and a systematic testing framework for computing crystalline material properties \cite{karls2020openkim}. We point the reader to the Supplemental Information (SI) for a detailed description of OpenKIM framework.

A key assumption in the proposed statistical inference approach is that correlations between properties discovered using IPs also apply to properties computed using first-principles methods (and by extension to real materials). 
In other words, we assume that the IP data and first-principles data belong to the same \textit{statistical pool}. 
If this is the case, then regression models developed based on IP data can be used to make first-principles predictions for large-scale QoIs using small-scale canonical properties computed from first principles. 
In addition, the properties determined by the regression model as strong predictors of the QoI (and associated atomic configurations) can be included in the training dataset of an IP being designed to model this QoI, thereby providing a systematic process for IP training.

To test this idea, we focus on symmetric tilt GBs in FCC metals.
A GB is an interface between two single crystals, each rotated differently.
The symmetric tilt GB energy depends on the interface plane (GB plane), and it is a function of the (symmetric) tilt angle of the two single crystals relative to this plane.
To capture the overall scaling of these GB energy curves, we adopt the lattice matching interatomic energy model (LM model) of Runnels et al.\ \cite{runnelsAnalyticalModelInterfacial2016}, which provides a functional form with a single material-dependent scaling factor $E_0$.
We use this scaling factor as the QoI for regression against a variety of canonical properties.
The training data for the regression model is obtained from the OpenKIM repository \cite{openkim:website}. 
Canonical properties and symmetric tilt GB energies across four tilt axes ([001], [111], [110], [112]) for ten FCC metals were computed for a subset of IPs in the OpenKIM repository using OpenKIM's automated testing workflow \cite{watersAutomatedDeterminationGrain2023}. 
The LM scaling factors computed from the GB energy calculations, along with the corresponding canonical property predictions, form the training set for a regression model.

GB energy is a good QoI for testing statistical inference for a number of reasons.
First, GB energy is simultaneously within reach of DFT calculations for some configurations but very costly for others depending on the tilt angle. 
For example, a supercell of a nickel symmetric tilt GB for the [110] tilt axis with a tilt angle of $109.47^\circ$ contains 46 atoms (computable via DFT), compared with 2586 atoms for a tilt angle of $77.89^\circ$ (generally out of scope for DFT). 
Second, a database of DFT simulated GBs exist as a source of ``ground truth'' reference data for the purposes of our study \cite{zhengGrainBoundaryProperties2020}.
Third, previous atomistic simulations of GB energies using IPs have demonstrated substantial variation across materials, GB types, and IPs \cite{wolfCorrelationEnergyStructure1989, wolfCorrelationEnergyStructure1990, wolfStructureenergyCorrelationGrain1989, wolfStructureenergyCorrelationGrain1989a, wolfStructureenergyCorrelationGrain1990, wolfStructureenergyCorrelationGrain1990a, olmstedSurveyComputedGrain2009, 
holmComparingGrainBoundary2010,
Tschopp2015,watersAutomatedDeterminationGrain2023}. 
This provides a statistically rich pool of data for training a regression model.\footnote{Given that a large portion of work in the field of GBs is based on atomistic simulations --- often without comparison to experiment, and only rarely with comparison to first-principles calculations due to the computational cost --- the wide spread in atomistic simulation findings is concerning. The results of our data-driven study can help to assess the validity of past simulations and help guide future work.}

The validity of our results depend on the reliability of the training set.
This is strengthened by several factors.
First, the overall quality of the IPs used to generate the training set is ensured by using IPs archived in OpenKIM IPs. All OpenKIM IPs are subjected to strict provenance tracking and \textit{verification checks} that ensure their correct implementation and satisfaction of basic physical invariances (such as objectivity). 
Second, the same computational protocol (OpenKIM test driver) \cite{watersAutomatedDeterminationGrain2023} was used to compute GB energies for all IPs. This avoids additional uncertainty due to the details of GB construction (we refer the reader to the cited reference for the specifics).
Third, most (if not all) of the IPs in the OpenKIM database were not fit to GB energies; therefore, each IP provides an independent prediction for the ``true'' material value.
Fourth, the use of a statistical approach addresses three additional concerns: 1) the inclusion of unsuitable models (e.g., IPs for compounds that were never meant to model elemental systems), 2) the variety of functional forms in physics-based IPs, each with its own biases, and 3) data-driven IPs that may be used outside their range of validity. If a reliable regression model is obtained, then it can be argued that the effect of such cases, if they exist, is of minimal importance.

Based on our training set, we construct a regression model with error bounds to predict the GB energy scaling factor from canonical properties. We find that the model exhibits excellent predictive ability. Further, we show that this regression model, fit on IP data, holds for DFT data as well, thereby illustrating that IP models and DFT models belong to the same statistical pool. In the process, we recover well-known relationships between various properties and GB energy \cite{wolfBrokenbondModelGrain1990, udlerGrainBoundarySurface1996, wynblattCorrelationGrainBoundary2001, suzukiInteractionPointDefects2003, wynblattRelationGrainBoundary2005, holmComparingGrainBoundary2010,  ratanaphanGrainBoundaryEnergies2015, zhangCoreShiftControls2023} and identify new ones. Further, the identified correlations indicate which canonical properties are important for fitting IPs that are to be used in applications where GB energetics are important (e.g., mechanical deformation of polycrystalline materials). This approach generalizes to the training of IPs for other QoIs.

The paper is structured as follows. 
Section~\ref{sec:sim-methods} describes the methods used including a brief review of symmetric tilt GBs and the LM model, data collection and analysis methods, and predictive model development.
Section~\ref{sec:results} presents the results including the input data analysis, regression model results, and DFT property predictions. Section~\ref{sec:discussion} provides a discussion of the observed correlations, predictive model performance, and the significance of specific property groups. The paper concludes in Section~\ref{sec:conclusion} with a summary and suggestions for future work.

\section{Methods}
\label{sec:sim-methods} 

We begin by describing the lattice-matching model for GB energy from which the desired energy scaling coefficient is derived.
This is followed by the data collection and analysis methods.
We close this section with a discussion of the support vector regression and multiple linear regression methods used in our analysis.

\subsection{Symmetric Tilt GBs and the Lattice-Matching Model} 
A tilt GB is constructed by (1) passing a plane (called the GB plane) through an infinite crystal, creating two half crystals and extending them past the GB plane so that their atoms overlap; (2) defining a tilt axis lying in the GB plane; (3) rotating each crystal by some angle around the tilt axis; and (4) deleting any atoms falling outside its original half-crystal domain. 
The tilt GB is \textit{symmetric} when the two crystals are rotated by opposite angles (i.e., one crystal is rotated by $\theta$ and the other by $-\theta$). 
The symmetric tilt GB energy is defined as the energy per unit area associated with the GB interface, and it is sensitively dependent on the tilt angle (e.g., see \cref{fig:gb_energy}). 

To describe GB energetics across a range of orientations, we adopt the LM model of Runnels el al.\ \cite{runnelsAnalyticalModelInterfacial2016, runnelsRelaxationMethodEnergy2016}. 
We provide a brief summary here, referring the reader to the detailed derivation and figures in the original references.
The LM model estimates GB energy based on bicrystals composed of rigid lattices, i.e., without accounting for relaxation of atomic positions.
This eliminates the need to resolve microscopic degrees of freedom, and allows the energy to be estimated analytically as a function of macroscopic degrees of freedom only. The GB between a ``black'' lattice $\mathcal{L}^\rb$ and a ``white'' lattice $\mathcal{L}^\rw$ (corresponding to the two sides of the GB plane) is represented in terms of the number density fields of the two lattices defined as functions of position in space $\mathbf{x}$:
\begin{equation}
    \rho^{\rb,\rw}(\mathbf{x}) \equiv \psi(\mathbf{x})\star
    \sum_{\mathbf{x}_n \in \mathcal{L}^{\rb,\rw}} \delta(\mathbf{x} - \mathbf{x}_n),\label{eq:rho}
\end{equation}
where $\star$ denotes the convolution operator, $\delta$ is the Dirac delta, and $\psi$ is a Gaussian thermalization function 
\begin{equation}
    \psi(\mathbf{x}) = \frac{1}{\sigma^3 \pi^{3/2}}e^{-\|\mathbf{x}\|^2/\sigma^2}.
\end{equation}
Physically, the standard deviation $\sigma$ can be understood as a thermalization parameter capturing the entropic effects of temperature. Alternatively, it can be interpreted as a surrogate for local elastic relaxation. However, in practice, it is taken to be a tuning parameter.

The GB energy $\gamma$ between the black and white lattices is given by the LM model as
\begin{equation}
    \gamma(\rho^\rb, \rho^\rw) = E_0 \left( 1 - \frac{c[\rho^\rb, \rho^\rw]}{\sqrt{c_0 [\rho^\rb] c_0 [\rho^\rw]}} \right),
    \label{eq:cov1}
\end{equation}
where $E_0$ [J/m$^2$] is a material dependent scaling factor, $c$ is the LM covariance of the densities, and $c_0$ is the ideal LM covariance assuming rotational transformations.\footnote{Note that this ``LM covariance'' is not the same as the statistical covariance between canonical properties and the scaling factor, discussed elsewhere in this manuscript.}  The LM covariance is defined as the normalized integral of the density function product across the GB plane \cite{runnelsRelaxationMethodEnergy2016}:
\begin{equation}
    c[\rho^\rb, \rho^\rw] \equiv \fint \rho^\rb(\mathbf{y}) \rho^\rw(\mathbf{y}) w_\lambda (\mathbf{y}) d\mathbf{y},
    \label{eq:cov}
\end{equation}
where the slashed integral denotes normalized integration over the two-dimensional GB plane, and $w_\lambda$ is a localization (or window) function with scale $\lambda$. For infinite, periodic lattices, the lattice densities in \cref{eq:rho} can be expressed as Fourier series, which enables the evaluation of \cref{eq:cov} in closed form.
It is apparent from \cref{eq:cov1} that the GB energy is bounded between $0$ and $E_0$.
In principle, $\gamma$ is a strict upper bound due to the lack of atomic relaxation.

Inputs to the LM model in \cref{eq:cov1} include the tilt axis, tilt angle, and crystal structure (specifically, the lattice constant).
The model requires three parameters: the localization function scale $\lambda$, a thermalization parameter $\sigma$, and the scaling factor $E_0$. 
It is the scaling factor $E_0$ that we are interested in predicting with our forward model, and the other two parameters will be taken from literature.

The GB energy can now be computed across a range of tilt axes and angles for various species, assuming the scaling factor is known.
In practice, the GB energy in \cref{eq:cov1} is modified to obtain a better upper bound estimate by accounting for facet relaxation that reduces the GB energy \cite{runnelsRelaxationMethodEnergy2016}.

\subsection{Data collection and analysis methods}
\label{sec:data_collection}
\begin{figure}
    \centering
    \includegraphics[width=16cm]{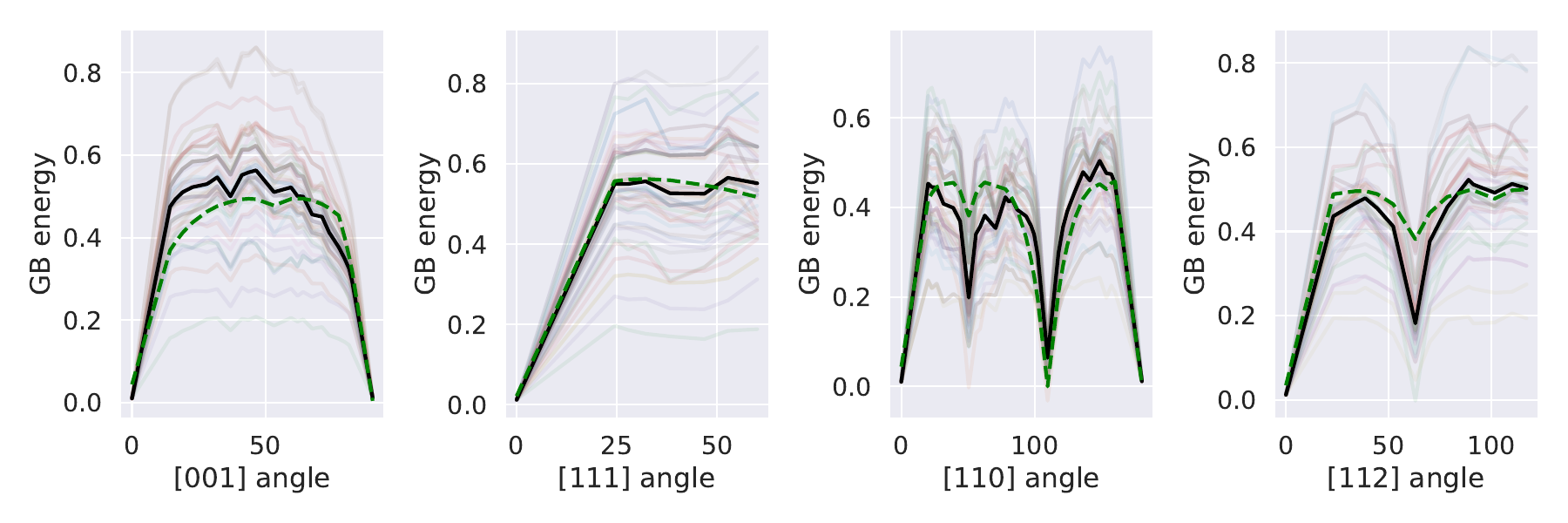}
    \caption{Sample simulation results for the GB energy [J/m\textsuperscript{2}] versus tilt angle for different symmetric tilt boundaries for IPs for aluminum drawn from the OpenKIM repository (additional species provided in the SI). See \cite{watersAutomatedDeterminationGrain2023} for model details. The average atomistic GB energy from OpenKIM (solid black) is compared with the average best fit analytical LM model (dashed green).
    } 
    \label{fig:gb_energy}
\end{figure}

At the time of writing, the OpenKIM repository contains the computed symmetric tilt GB energies from 304 IPs \cite{watersAutomatedDeterminationGrain2023}.
As an example, \cref{fig:gb_energy} shows results for aluminum IPs in OpenKIM  for the GB energy versus tilt angle for different symmetric tilt boundaries (referred to as ``GB energy curves'' from here  on).
A total of 1590 GB energy curves were extracted from OpenKIM through queries to the public facing database, of which 1269 are from FCC structures.
153 results were removed whose predicted GB energies are below -0.1 or greater than 100 J/m$^2$, which may indicate calculation failures.
For example, IPs that do not have FCC as a ground state may experience a relaxation to the body-centered cubic (BCC) structure, or exhibit alternating twins within the bulk away from the GB.
Finally, 76 pair potential results were removed from the analysis pool, as they are lacking the many-body density-dependent terms necessary for metallic materials \cite{vitekPairPotentialsAtomistic1996}.
This leaves 1040 GB energy curves across four tilt axes: [001], [110], [111], [112]. 

\begin{table}[]
    \centering
    \caption{Lattice constants $a$ used for the analytical LM model.}
    \begin{tabular}{lllllllllll}
    \hline
         & Ag & Al & Au & Cu & Fe & Ni & Pb & Pd & Pt & Rh \\ \hline
        $a [\text{\AA}]$ & 4.082 & 4.046 & 4.065 & 3.615 & 2.856 & 3.499 & 4.95 & 3.89 & 3.92 & 3.9 \\
    \hline
    \end{tabular}
    \label{tab:matl_params}
\end{table}

The empirically determined scaling coefficient ($E_0$) for the retained GB energy curves requires that we calculate the unscaled GB energy LM model predictions (i.e., \cref{eq:cov1} with $E_0$ set to 1) for 40 unique cases (4 tilt axis $\times$ 10 species).
These unscaled results are computed using the \texttt{wield} code, available on Github \cite{runnelsSolidsgroupWIELDRepository}. The non-dimensional temperature parameter ($\sigma/\alpha$) is set to 0.175 and the window scaling parameter ($\lambda$) is set to 0.5, determined to be suitable for FCC structures in previous work \cite{runnelsbrandonscottModelEnergyMorphology2015, runnelsAnalyticalModelInterfacial2016}. The lattice constants used in the LM model calculations, given in \cref{tab:matl_params}, are taken as an average over the IP predictions extracted from OpenKIM (removing outliers). 

The scaling factor $E_0$ for each IP and species combination is calculated using a least-squares approach across all four tilt axes.
This calculation across tilt axes leads to the final dataset of 300 scaling coefficients from which we work.
As a demonstration, the analytical LM model for four tilt axes with aluminum are shown in \cref{fig:gb_energy}. 
The average of the GB energy simulation results is compared with the analytical model using the average scaling coefficient.
This average scaling coefficient is only used for demonstration purposes here and is calculated across all 235 unique IP models that were used. 
(See the SI for the references of all IPs used and a breakdown by model type.)

\begin{table}[t]
\small
\centering
\caption{Canonical properties and associated crystal structures (body-centered cubic (BCC), face-centered cubic (FCC), simple cubic (SC)) extracted from OpenKIM, along with the property abbreviations used in this paper. For example, the cohesive energy, abbreviated as ``CohEn'' in figures in this paper, for the equilibrium crystal structures of the BCC, FCC and SC structures. $^*$Note that the bulk modulus is computed from the elastic constants, so it is not an independent property.
}
\begin{tabular}{lll}
\hline
Canonical Property                               & Structures   & Property Abbreviation \\ \hline
Bulk Modulus$^*$                                     & BCC, FCC, SC & Bulk Mod.             \\
Cohesive energy                                & BCC, FCC, SC & CohEn                 \\
Elastic constants ($C_{11}$, $C_{12}$, $C_{44}$)                                    & BCC, FCC, SC & C11, C12, C44         \\
Extrinsic stacking fault energy                  & FCC          & eSFE                  \\
Intrinsic stacking fault energy                  & FCC          & iSFE                  \\
Lattice constant                                 & BCC, FCC, SC & LC                    \\
Relaxed vacancy formation potential energy       & BCC, FCC     & rVFPE                 \\
Surface energy (\{100\} \{110\} \{111\} \{112\})   & BCC, FCC     & SE                    \\
Thermal expansion coefficient                    & BCC, FCC     & TEC                   \\
Unstable stacking fault energy                   & FCC          & uSFE                  \\
Unstable stacking fault energy slip fraction     & FCC          & uSFESF                \\
Unstable twinning energy                   & FCC          & uTwE                  \\
Unstable twinning energy slip fraction     & FCC          & uTwESF                \\
Unrelaxed vacancy formation potential energy     & BCC, FCC     & uVFPE                 \\
Vacancy migration energy                         & BCC, FCC     & VME        
                   \\
Vacancy relaxation volume                                & BCC, FCC     & VRV \\
\hline
\end{tabular}
\label{tab:props}
\end{table}

A set of canonical properties associated with the 235 retained IPs is extracted from OpenKIM (see \cref{tab:props}).
Each computed scaling factor $E_0$ is paired with its respective canonical properties on the basis of IP and species to form the final database used in our analysis.
Initial inspection of each canonical property distribution revealed a few values that were significantly outside what would be considered reasonable values.
The existence of these outliers may be due to any number of reasons, including improper fitting of the IP or property prediction in crystal structures that the IP was not intended.
To prevent these outliers from having an undue influence on the analysis, a set of property cutoffs were imposed on the canonical properties (see the SI for box plots justifying the cutoffs, along with their values).

It is important to identify which properties are correlated with each other, as well as with GB energy.
This pairwise correlation between properties is determined using Pearson's correlation coefficient, defined for two variables $(x,y)$ as:
\begin{equation}
    \label{eqn:corr_coeff}
    r_{x,y} = 
    \frac{\sum_{i=1}^N (x_i - \bar{x})(y_i - \bar{y})}{\sqrt{\sum_{i=1}^N (x_i-\bar{x})^2} \sqrt{\sum_{i=1}^N (y_i-\bar{y})^2}},
\end{equation}
where $\bar{x}$ and $\bar{y}$ are the average values over $N$ observations of the variables $(x_i,y_i)$ ($i=1,\dots,N$).
Highly correlated variables will have a value approaching $1$ (positive correlation) or $-1$ (negative correlation).

\subsection{Data-driven canonical-property-to-scaling-coefficient regression model}
\label{sec:model}
We now move to the development of a regression model for predicting the GB energy scaling coefficient from canonical properties.
To begin, we determine the mean and variance of the canonical property data extracted from OpenKIM.
This is used to scale all data to a mean of 0 and variance of 1.
While the relevant canonical properties are typically calculated at time of upload to the OpenKIM database, there are instances where IP models are missing certain property values.
A $k$-nearest neighbor (KNN) imputer is leveraged to allow the use of IPs with missing values (KNNImputer in Scikit-Learn).
This method finds the $k$-nearest neighbors (we use $k=2$) using a Euclidean distance metric that can accommodate the missing coordinate values. 
This is accomplished by taking the average of the squared difference for values which \textit{are} present, and multiplying the result by the number of dimensions (e.g. the number of possible canonical properties used in the model) (see \cite{SklearnMetricsPairwise2024} for example). 

There are a variety of modeling approaches to choose from when developing the forward model. 
In this work, we utilize two approaches available in the Scikit-Learn package \cite{pedregosaScikitlearnMachineLearning2011}.
The first is support vector regression (SVR), an adaptation of support vector machines for regression problems \cite{smolaTutorialSupportVector2004}.
This approach uses a set of hyperplanes and a non-linear radial basis function ($\exp(-\gamma \| x - x'\|^2)$) as the kernel \cite{smolaTutorialSupportVector2004}.
In addition, we fit a multilinear regression model that, due to its simpler form, is easier to understand.

We leveraged as many canonical properties as possible from the OpenKIM database in developing our dataset.
To help elucidate which properties have the largest influence on GB energy, we explore permutations of three-factor models and their predictive performance.

We use $k$-fold cross-validation (CV) tests with all possible permutations of up-to-three-factor models to select the best one to report.
$k$-fold CV is a well-established technique for evaluating model performance with limited data.
To start, three canonical properties are selected for evaluation.
The dataset is divided into $k$ subsets called ``folds.''
Each fold is used once as the test data, while a model is fit or trained on the remaining folds.
This is repeated across all $k$ folds, and then repeated across the data a number of times by defining new folds.
We then take the average of the root mean square error across all repeated folds as the performance metric for that specific combination of factors.
The best models result in the smallest root mean square error.
In this work, we use $k=10$ folds, repeated 3 times.

SVR models have additional hyperparameters ($C$ and $\epsilon$) that must be determined to give the best performing model.
This is built into the process described above using nested $k$-fold CV.
The data is still divided into $k$ folds, with a unique set of hyperparameters selected for fold through an additional inner $k$-fold CV study. 
The outer CV provides the estimate for the overall model error for the given properties, while the inner CV determines the optimal hyperparameters for the current test fold.

\subsection{First-principles calculations}
\label{sec:dft}
We leverage existing DFT grain boundary energy results from literature \cite{zhengGrainBoundaryProperties2020}.
These data are used to obtain the DFT GB energy scaling coefficients, for comparison with the regressed predictions.
We point the reader to the literature \cite{zhengGrainBoundaryProperties2020} for details, but we note here that the GB energy calculations were performed using the Perdew-Berke-Ernzerhof (PBE) generalized gradient approximation (GGA) functional.
For consistency, our DFT canonical property calculations make use of the same functional.
The DFT calculations of predictors follow the methods outlined in \cite{jaspersonCrossscaleCovarianceMaterial2024} with the functional change noted above (PBEsol to PBE).

\section{Results}
\label{sec:results}

\subsection{Input data analysis}
\begin{figure}[t]
    \centering
    \includegraphics{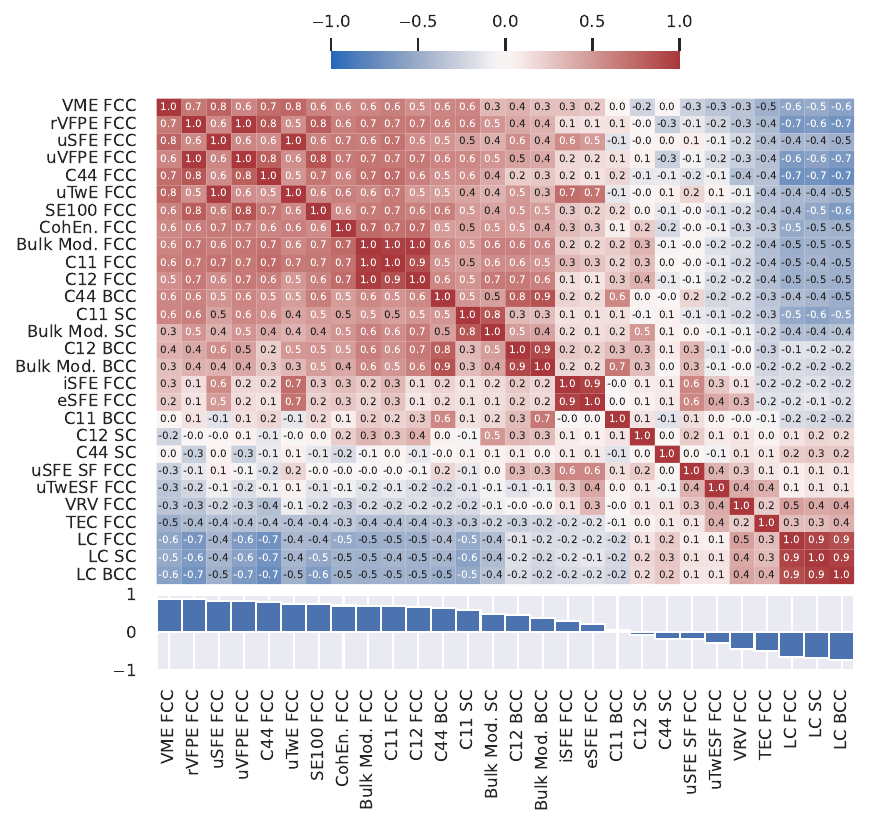}
    \caption{Heatmap of Pearson correlation coefficients between canonical properties (see \cref{tab:props} for the abbreviations), with the bar chart showing correlation coefficient between the $E_0$ GB energy scaling coefficient and canonical properties. 
    \label{fig:corr_heatmap_manuscript}
    }
\end{figure}

\begin{figure}
    \centering
    \includegraphics[width=0.8\textwidth]{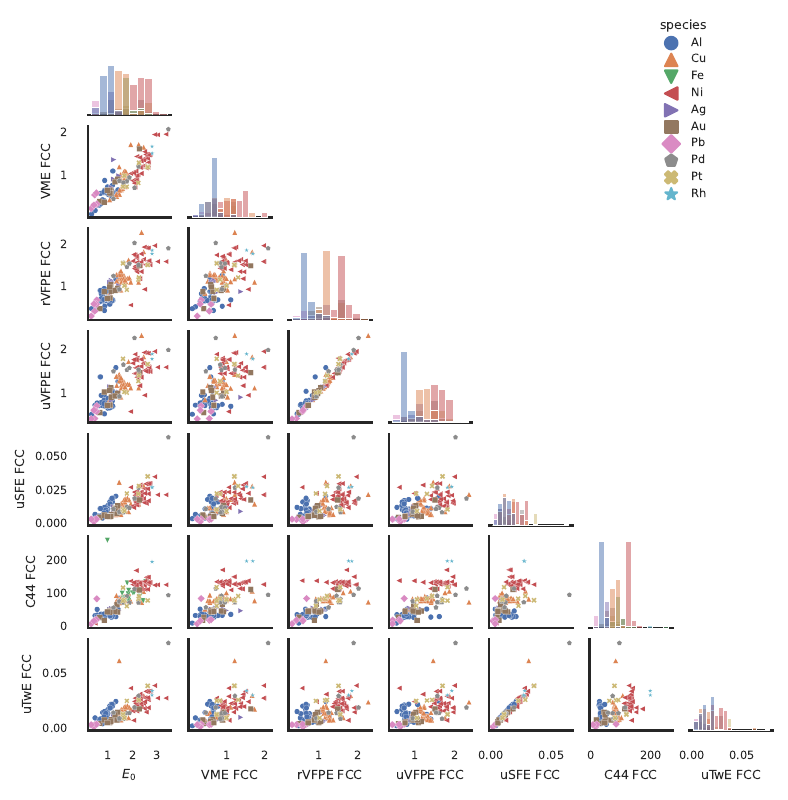}
    \caption{Pairwise relationships (pairplots) of the canonical properties most correlated with the GB energy scaling coefficient, $E_0$ [J/m\textsuperscript{2}], including the vacancy migration energy [eV], relaxed and unrelaxed vacancy formation potential energies [eV], unstable stacking fault energy [eV/A\textsuperscript{2}], and $C_{44}$ elastic constant [GPa]. Off-diagonal plots present the scatter plots of the different variable combinations, with a histogram for each property along the diagonal.}
    \label{fig:pp_top}
\end{figure}

The full set of data and calculated GB scaling coefficient $E_0$ are provided in Github \cite{jaspersonGBCovarianceGitHub2024}.
The heat map in \cref{fig:corr_heatmap_manuscript} presents the measured pairwise correlations for all canonical properties in \cref{tab:props}.
A bar chart below the heat map displays the calculated correlation coefficient for each property with the scaling coefficient $E_0$. 
To facilitate the analysis and discussion, a number of pairplots were produced with various canonical properties and $E_0$ (\cref{fig:pp_top}).
The off-diagonal plots show scatter plots of the different variable combinations.
These correlations are discussed in detail in \cref{sec:discussion_properties}.

\subsection{Modeling results}
\begin{figure}[t]
    \centering
    \includegraphics[width=14cm]{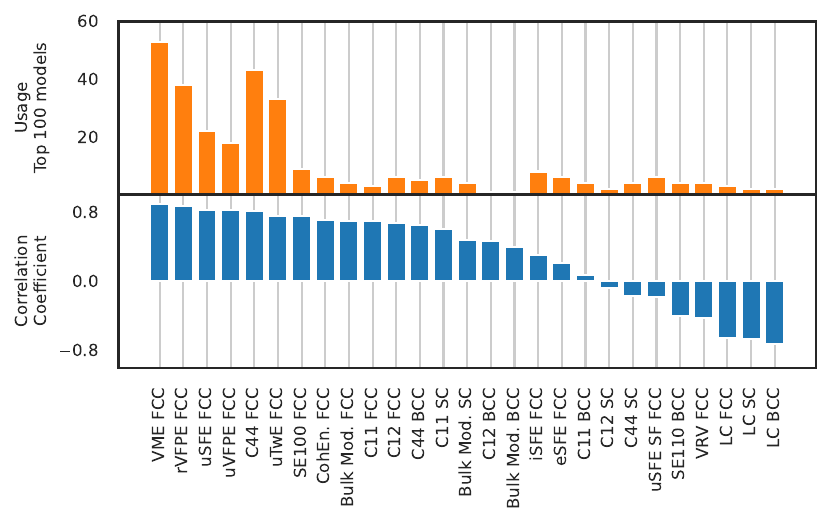}
    \caption{Factor influence assessment. 
    Orange bars display the factor prevalence in the top 100 models, sorted using the average root mean square error across folds from $k$-fold cross-validation. 
    Blue bars display the correlation coefficient between the factor and the GB energy scaling coefficient.
    }
    \label{fig:factor_influence}
\end{figure}

We report the factor selection results from the $k$-fold CV study.
The top 5 performing models and their predictor properties are listed in \cref{tab:top5_models}.
Given that the top models all have similar performance, it is helpful to examine a histogram of factor usage in the top 100 models.
\Cref{fig:factor_influence} shows the prevalence of each factor, along with its correlation to the $E_0$ GB energy scaling coefficient.

\begin{table}[]
\centering
\caption{The three factors leveraged in the top 5 SVR models, as evaluated using the average RMSE across all test folds from the repeated $k$-fold cross-validation study. uTwE, uSFE and VME are highly cross-correlated with each other, making them interchangeable in terms of model performance.}
\begin{tabular}{llll}
\hline
Factor 1 & Factor 2 & Factor 3 & \begin{tabular}[c]{@{}l@{}}CV Score \\ (avg. RMSE)\end{tabular} \\ \hline

uSFE FCC	& rVFPE FCC	& VME FCC	& 0.185 \\
uTwE FCC	& rVFPE FCC	& VME FCC	& 0.185 \\ 
uTwE FCC	& uVFPE FCC	& VME FCC	& 0.189 \\
C11 SC	    & C44 FCC	& uTwE FCC	& 0.194 \\
C12 FCC	    & C44 FCC	& VME FCC	& 0.196 \\
     \hline

\end{tabular}

    \label{tab:top5_models}
\end{table}

\begin{figure}
    \centering
    \subfloat[\centering]{{\includegraphics{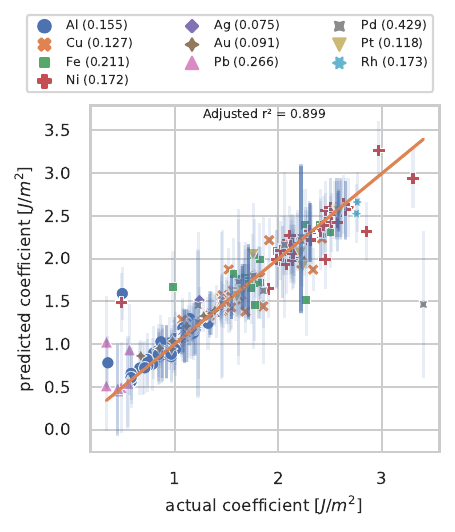}}}
    \subfloat[\centering]{{\includegraphics{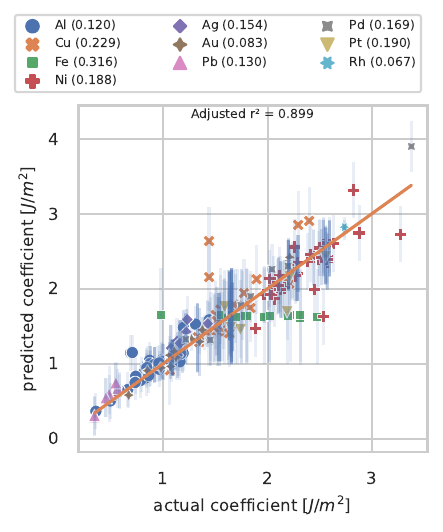}}}
    \caption{Regression model prediction versus actual GB energy scaling coefficients, as predicted by models using nested $k$-fold CV for SVR hyperparameter selection and root mean squared error (RMSE) estimate using (a) all canonical properties, and (b) 3-factor (uSFE, rVFPE, VME) model using linear regression, see \cref{eqn:linear_3factor}. The root mean squared error (RMSE) as a function of species is given in the legend. Error bars represent estimate for prediction uncertainty (1.96x RMSE) for each species.
    }
    \label{fig:nested_cv_uncert}
\end{figure}

Using the top performing three-factor model from an analogous $k$-fold study, the three-factor multilinear regression model, without scaling, is given by: 
\begin{align}
    \notag
    \text{coeff} = 
    18.52125 \times \text{uSFE-FCC} 
    + 
    0.79610 \times \text{rVFPE-FCC} 
    + \\
    \label{eqn:linear_3factor}
    0.67369 \times \text{VME-FCC} 
    - 
    0.15919.
\end{align}

Finally, evaluation of the model using nested (for hyperparameter optimization and RMSE estimate) $k$-fold CV is shown in \cref{fig:nested_cv_uncert}.

\subsection{DFT properties and regression prediction}
\begin{figure}[t]
    \centering
    \includegraphics{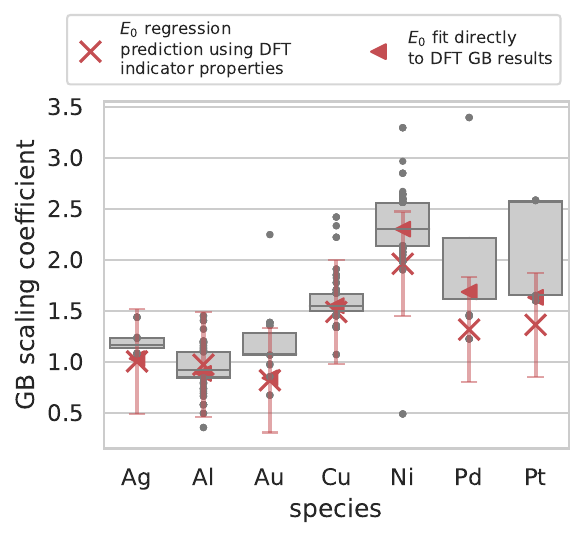}
    \caption{Comparison of calculated GB energy scaling coefficients [J/m\textsuperscript{2}]. Red $\bigtimes$ shows the predicted scaling coefficient using a linear regression model fit using three predictors: $C_{44}$ elastic constant, intrinsic stacking fault energy, and unstable stacking fault energy (all FCC). Red triangle shows calculated scaling coefficient from a direct fit against DFT GB energy results. Gray boxes show quartiles of calculated scaling coefficient values from the IP model dataset. Gray circles indicate calculated scaling coefficient results that are outside of the quartiles.
    }
    \label{fig:dft}
\end{figure}

We filter the nested $k$-fold CV evaluation to only use indicator properties for which we have available DFT calculated values (LC, C44, SE111, uSFE, iSFE) and select the top performing model (C44, iSFE, uSFE).
We then fit the three-factor regression model using the IP data for these three properties and GB energy. 
Next, using the DFT calculated canonical properties, we make a prediction of the GB energy scaling coefficient from first-principles data.
This regressed coefficient is compared with coefficients calculated directly from first-principles GB energy data (\cref{tab:dft}).
\Cref{fig:dft} presents the regressed-on-DFT-indicator scaling coefficient in comparison with the direct fit scaling coefficient.
Boxplots of the scaling coefficients calculated using all available IP models in OpenKIM are provided for reference.

\begin{table}[]
\centering
\caption{Comparison of calculated scaling coefficient using DFT GB energies versus the predicted value using a regression model with DFT canonical properties. The calculated values for the three indicator properties made using DFT with the PBE functional are shown. See \cref{sec:dft} for the simulation details.}
\label{tab:dft}
\begin{tabular}{lllllll}
\hline
Species & \begin{tabular}[c]{@{}l@{}}C44\\ FCC\\ (GPa)\end{tabular} & \begin{tabular}[c]{@{}l@{}}iSFE\\ FCC\\ (meV/A\textsuperscript{2})\end{tabular} & \begin{tabular}[c]{@{}l@{}}uSFE\\ FCC\\ (meV/A\textsuperscript{2})\end{tabular} & \begin{tabular}[c]{@{}l@{}}Coefficient,\\ exact fit using\\ DFT GB\\ energies\end{tabular} & \begin{tabular}[c]{@{}l@{}}Coefficient,\\ regressed fit \\ using DFT \\ indicator\\ properties\end{tabular} & \begin{tabular}[c]{@{}l@{}}Percent\\ Error\end{tabular} \\ \hline
Ag & 41.208 & 1.067 & 6.472 & 1.032 & 1.004 & -2.68 \\
Al & 33.660 & 7.602 & 10.430 & 0.886 & 0.972 & 9.73 \\
Au & 26.529 & 1.872 & 5.418 & 0.842 & 0.820 & -2.67 \\
Cu & 79.597 & 2.628 & 11.104 & 1.551 & 1.490 & -3.93 \\
Ni & 117.592 & 8.919 & 17.851 & 2.302 & 1.963 & -14.76 \\
Pd & 59.541 & 9.100 & 14.168 & 1.686 & 1.317 & -21.91 \\
Pt & 60.493 & 18.993 & 19.973 & 1.633 & 1.363 & -16.5 \\ \hline
\end{tabular}
\end{table}

\section{Discussion}
\label{sec:discussion}

\subsection{Correlation between canonical properties}
We begin by examining the pairwise correlation coefficients, defined in \cref{eqn:corr_coeff} and presented in \cref{fig:corr_heatmap_manuscript}. 
FCC properties that are highly correlated with the scaling coefficient include vacancy migration energy (VME), relaxed and unrelaxed vacancy formation potential energy (rVFPE and uVFPE, respectively), unstable stacking fault energy (uSFE), and the $C_{44}$ elastic constant, among others (these correlations with GB energy are discussed further in \cref{sec:discussion_properties}).

Many canonical properties are strongly correlated with each other.
This is evident in \cref{fig:corr_heatmap_manuscript} by the uniform-colored square regions along the diagonal for the vacancy formation potential energies, lattice constants, and others (surface energies are not shown here but can be seen in the SI).
Pairplots of these highly correlated properties are included in the SI as additional justification.
For the purposes of property influence, it seems sufficient to consider one property from each of the following highly correlated property groups: bulk modulus/$C_{11}$/$C_{12}$, surface energies, uVFPE/rVFPE, lattice constants, iSFE/eSFE, and uSFE/uTwE.

Two model architectures are shown in \cref{fig:nested_cv_uncert}: all factor SVR and three-factor multilinear regression (uSFE, rVFPE, VME).
Both models perform similarly with excellent prediction capabilities and reasonable errors, as illustrated by $r^2$ values around 0.9 and the reported RMSE for each species (shown in the legend).
Given the comparable performance of these two models, we focus on the simpler 3-factor multilinear model from here on.

\subsection{Specific property group significance and their relation to grain boundary energy}
\label{sec:discussion_properties}

While the observed correlations between properties are informative, they cannot identify causal relationships between the canonical properties and GB energy.
In this section, we present the most significant factors and their appearance in existing models in literature. 
We highlight this information as supporting evidence that known physical trends are indeed maintained across the IPs and data, strengthening the argument that they serve as valid ``synthetic materials.'' 

To identify significant canonical properties, we look at both the pairwise correlation coefficient, \cref{eqn:corr_coeff}, and the factor prevalence in the top 100 models, as identified using $k$-fold CV (\cref{fig:factor_influence}, sorted by decreasing correlation coefficient).
As a reminder, a strong correlation with GB energy is observed on both ends of the histogram, with correlations approaching $1$ or $-1$.
We note that vacancy migration energy (VME) is the most used factor.
This is perhaps not surprising, given that it is the most correlated factor with the scaling coefficient.
The next two factors, relaxed and unrelaxed vacancy formation potential energy, are strongly correlated with each other.
Combining these together under the vacancy formation potential energy (VFPE) heading results in a subgroup that is used almost as frequently as VME.

\Cref{fig:pp_top} shows a strong correlation between the GB energy scaling coefficient, the VME, and the VFPE.
GB energy and its relation to vacancy formation and diffusion have been the subject of much research over the past few decades (see \cite{herzigGrainBoundaryDiffusion2005}, for example).
It is well-known that locally, vacancy formation and diffusion within a GB are strongly influenced by the large interface stresses.
This results in localized property variations throughout the GB, with generally lower VFPE, lower VME, and larger diffusion coefficients as compared to the bulk.
Furthermore, the connection between GB energy and the lowest VFPE present within the GB has previously been shown \cite{suzukiDiffusionMechanismsGrain2004}. 
What we show instead is how the overall GB energy corresponds with the bulk material VME and VFPE.
The $E_0$ scaling factor from the LM model is a single scalar value characterizing the material dependence of the entire GB-energy-versus-tilt-angle symmetric tilt GB energy curve. 
Since the VME and VFPE are calculated in the bulk for each IP, it is possible to relate bulk properties to this scalar.
The data shows that the GB energy increases as it becomes harder to form vacancies in the bulk.
It is important to note that the positive correlation between GB energy and VME holds across IPs and species (\cref{fig:pp_top}, first row, first column).
This highlights how ``synthetic materials'' can help elucidate property correlations seen between materials by providing additional data beyond that provided by actual materials.

\begin{figure}[t]
    \centering
    \includegraphics[width=0.8\textwidth]{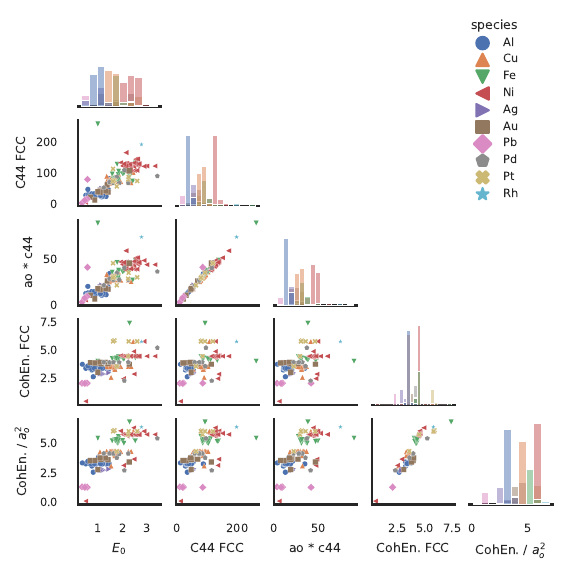}
    \caption{
        Scatter plots highlighting different combinations of GB energy scaling coefficient, $E_0$, and select canonical properties calculated from IPs, with a histogram of available IP data for each property along the diagonal. A strong linear relationship for $E_0$ is observed with $C_{44}$ and, to a lesser extent, cohesive energy. We observe with CV testing that scaling $C_{44}$ by $a_0$ and the cohesive energy by $1/a_0^2$ does not result in a more predictive model. $E_0$, CohEn$/a_0^2$ and $a_0 \times C_{44}$ are in J/\,m\textsuperscript{2}, bulk modulus and $C_{44}$ are in GPa, CohEn is in eV, and lattice constant is in Angstroms.
        }
    \label{fig:pp_specific}
\end{figure}
After vacancy-related factors and SFE, the property most correlated with the GB energy scaling coefficient is the $C_{44}$ elastic constant (shear modulus), shown in \cref{fig:pp_specific}.
The relationship between GB energy and $C_{44}$ has been discussed in the literature for some time \cite{udlerGrainBoundarySurface1996}.
It is often combined with the lattice constant to reflect how GB energies scale between materials \cite{holmComparingGrainBoundary2010, suzukiInteractionPointDefects2003, zhangCoreShiftControls2023}.
This correlation makes sense, given that a dislocation's elastic energy is proportional to the shear modulus \cite{readDislocationModelsCrystal1950,hullIntroductionDislocations2011}. 
We note that, despite its strong correlation with $E_0$ and its frequency of use in the top 100 models, $C_{44}$ is outperformed in the top performing models by the stacking fault energies and vacancy formation/migration energies.
This is likely due, at least in part, to the strong covariance between the canonical properties themselves (see \Cref{fig:pp_top}).

\Cref{fig:pp_specific} also shows the relation between the cohesive energies and the scaling coefficient.
A GB interface disrupts the bonds between atoms \cite{ratanaphanGrainBoundaryEnergies2015}, which is the basis for broken bond models for GB energy \cite{wolfBrokenbondModelGrain1990,wynblattCorrelationGrainBoundary2001, wynblattRelationGrainBoundary2005}.
In past work, the cohesive energy scaled by the lattice constant ($E_{\rm coh}/a_0^2$), has been shown to correlate with GB energy in BCC metals \cite{ratanaphanGrainBoundaryEnergies2015}.
While a linear relationship is predicted by the broken bond model for GB energy, it has been shown that $C_{44}$ is much more strongly correlated with GB energy in FCC metals \cite{suzukiInteractionPointDefects2003}.
This matches our observations in that the cohesive energy is not used in many of the top performing models. 
Additionally, cohesive energy can be roughly viewed as a scaling constant for the potential energy landscape, and is therefore implicitly included in every energy property (uSFE, VME, etc.).

As mentioned above, elastic constants and cohesive energy are often scaled using the lattice constant $a_0$ when exploring relationships to GB energy.
Dimensional analysis suggests that these canonical properties may be grouped into new variables using a fundamental length scale $\ell_0$, for instance, $\ell_0 C_{44}$ and $E_{\rm coh} / \ell_0^2$.
One natural choice of length scale is the lattice constant $a_0$; (see \cref{fig:pp_specific}); however this substitution does not induce any additional correlation.
Other more physical choices may be tied to length scales related to the GB structure, such as the range of influence of the GB in the direction normal to it. This is a question of interest for future work.

\subsection{Regression prediction from DFT indicator properties}
In \cref{sec:intro}, we made the assertion that IPs developed by experts to model specific material classes constitute an ensemble of synthetic materials whose predictions belong to the same statistical pool as those of first-principles calculations. 
We test this conjecture by examining whether our statistical model, developed on IPs, also accurately predicts DFT GB energies from DFT canonical properties.
We note from \cref{fig:dft} that all DFT predicted values fall within the error bars from the regression prediction, with reasonable errors reported in \cref{tab:dft}.
We therefore argue that the covariance between properties inferred from the IP results also applies to DFT properties and hence is a reflection of material behavior.

To understand why we believe this to be a valid hypothesis, we consider the IP's potential energy surface (PES), i.e., the potential energy as a function of atomic configuration.
The act of fitting an IP is an attempt to match a model's many-dimensional PES to that predicted from first principles.
A precisely matched PES will lead to accurate target property predictions.
We postulate that GB energy predictions require sampling regions of the model's PES that are also accessed during calculation of canonical properties strongly correlated with the GB energy scaling factor. 

Generalizing beyond this, we contend that an IP intended for modeling certain large-scale QoIs should be trained to reproduce canonical properties correlated with these QoIs. This validates the traditional view that successfully fitting an IP involves the ``art'' of selecting suitable properties to include in the training set \cite{brenner:2000}. 
The systematic identification of canonical properties correlated with a QoI through regression analysis, as done in this paper, can help improve the prediction accuracy of an IP by aiding in the fitting process.

As mentioned above, symmetric tilt GB energy predictions fall in the computational ``sweet spot'' for this research.
Current resources allow high symmetry tilt GBs to be calculated using first-principles methods \cite{zhengGrainBoundaryProperties2020}.
However, the number of GBs that meet these criteria are limited, reducing the available data.
IP-generated GB energy data is notably more comprehensive, allowing exploration of correlations with statistically significant sample sizes, with DFT data serving as a validation set for the results.
The advantage is only amplified when considering that the vast majority of large-scale QoIs are beyond the scope of first-principles calculations.
As an example, the authors have applied this approach to plastic flow strength simulations involving tens of millions of atoms, well beyond the reach of DFT \cite{jaspersonCrossscaleCovarianceMaterial2024}.
Certainly, the IP fitting process for these large-scale QoIs may benefit from the data-driven workflow presented here, while other ``approachable'' properties such as GB energy may not. We leave it to future research to explore this further.

\section{Conclusions}
\label{sec:conclusion}

We have shown how data from atomistic simulations using classical IPs can serve as an ensemble of ``synthetic materials'' for purposes of exploring the underlying physical mechanisms in real materials.
This approach allows us to uncover relations between fundamental microscopic properties computable from first principles (\textit{canonical properties}) and large-scale QoIs beyond the reach of first principles.
Central to the approach is the assumption that the synthetic materials and real materials belong to the same statistical pool, i.e., that correlations identified from IP calculations also apply to first-principles methods (and by extension to real materials).

As a proof of principle, we focus on symmetric tilt GB energies in FCC metals as the QoI. We chose this property due to the availability of first-principles results \cite{zhengGrainBoundaryProperties2020} for validation purposes, along with a high-quality computational protocol (test driver) in OpenKIM \cite{watersAutomatedDeterminationGrain2023}.
Exploration of the top performing models uncover the most used factors (canonical properties) that we infer to be significant predictors of GB energy: vacancy formation energy, vacancy migration energy, and the $C_{44}$ elastic constant.
We relate these factors to existing modeling approaches for GBs, such as arrays of dislocations and broken-bond models.

The approach presented here can be adopted for training IPs for QoIs for which first-principles results are not available: 1) Perform simulations with an ensemble of existing IPs to compute the QoI; 2) Develop a regression model relating the predicted QoIs to a pool of candidate canonical properties; 3) Incorporate first-principles values (and associated atomic configurations) of properties most correlated with the QoI into the training set of the new IP designed for simulating the QoI. 
An example of this approach applied to the plastic flow strength of metals, well beyond the reach of first-principles calculations, is described in \cite{jaspersonCrossscaleCovarianceMaterial2024}. Note also that the regression model can be used with first-principles values of the predictor properties to provide a direct estimate for the QoI.

There remain opportunities to strengthen the validation of our ``IPs as synthetic materials'' hypothesis by extending our dataset beyond symmetric tilt GBs to include more complex asymmetric and twist boundaries, and the addition of more factors into the canonical properties pool; this would involve the development of new OpenKIM test drivers for these cases.
Furthermore, opportunities may exist to leverage more sophisticated statistical inference methods, such as symbolic regression \cite{Schmidt2009S-Distilling} 
to uncover more fundamental relations. 

\section*{CRediT authorship contribution statement}
\textbf{Benjamin A. Jasperson}: Conceptualization, Data curation, Formal analysis, Investigation, Methodology, Resources, Software, Visualization, Writing -- original draft, Writing -- review \& editing. 
\textbf{Ilia Nikiforov}: Conceptualization, Formal analysis, Investigation, Methodology, Project administration, Resources, Supervision, Writing -- original draft, Writing -- review \& editing.
\textbf{Amit Samanta}: Formal analysis, Investigation, Resources, Software, Writing -- original draft, Writing -- review \& editing.
\textbf{Brandon Runnels}: Methodology, Software, Visualization, Writing -- original draft, Writing -- review \& editing.
\textbf{Harley T. Johnson}: Conceptualization, Funding acquisition, Investigation, Methodology, Project administration, Resources, Supervision, Writing -- review \& editing.
\textbf{Ellad B. Tadmor}: Conceptualization, Funding acquisition, Methodology, Project administration, Resources, Supervision, Writing -- original draft, Writing -- review \& editing.

\section*{Data Availability}
The data, Python code and implemented scikit-learn models will be made available at \cite{jaspersonGBCovarianceGitHub2024}: 
\begin{center}
    \url{https://github.com/Johnson-Research-Group/gb_covariance}
\end{center}

\section*{Acknowledgement}
The authors appreciate the support of Brendon Waters (UMN) in developing and implementing the grain boundary energy tests within the OpenKIM framework. 
This material is based in part upon work supported by the National Science Foundation under Grants No.\ 1834251, 1922758, and 2341922.
A part of this work was performed under the auspices of the U.S. Department of Energy at Lawrence Livermore National Laboratory under Contract DE-AC52-07NA27344 and was funded by the Laboratory Directed Research and Development Program under project tracking code 23-S1-006. The LLNL IM release number is LLNL-JRNL-867022.


\begin{singlespace}
	\printbibliography[heading=subbibintoc]
\end{singlespace}

\end{refsection}


\newpage
\section*{Supplementary Information for ``Fundamental Microscopic Properties as Predictors of Large-Scale Quantities of Interest: Validation through Grain Boundary Energy Trends''}
  
\newcommand{\beginsupplement}{%
	\setcounter{table}{0}
	\renewcommand{\thetable}{S\arabic{table}}%
	\setcounter{figure}{0}
	\renewcommand{\thefigure}{S\arabic{figure}}%
	\setcounter{section}{0}
	\renewcommand{\thesection}{S\arabic{section}}%
}
\beginsupplement 
\begin{refcontext}{supp}

\section{OpenKIM Framework}
\label{sec:openkim}

The Open Knowledgebase of Interatomic Models (OpenKIM) \cite{tadmor:elliott:2011,tadmor2013nsf} is a multifaceted cyberinfrastructure project founded in 2009 hosted at \url{https://openkim.org}. The components of OpenKIM enabling the present work are the IP repository and the automatic property testing framework \cite{karls2020openkim, karls:clark:2022}. At the time of submission, OpenKIM contains 643 curated IPs, each of which is a self-contained package compliant with the KIM API that allows it to be used in plug-and-play fashion with compatible simulation codes, such as ASE \cite{ase}, GULP \cite{gulp} and LAMMPS \cite{lammps}.


IPs archived in OpenKIM are identified by unique identifiers (KIM IDs).
Each IP is run through a battery of \emph{verification checks} (VCs) to ensure the correctness of an IP's coding, along with its conformance to basic physical requirements (e.g.\ invariances and smoothness).
Furthermore, \emph{tests} are run to compute an IPs prediction for a variety of material properties. 
These calculations are performed using an automated cloud-based workflow manager called the OpenKIM Processing Pipeline \cite{karls2020openkim} that matches IPs to compatible calculations (e.g.\ verifying that an IP supports the species required by a test), accounts for dependencies between tests, and runs the calculations on available HPC resources \cite{karls:clark:2022}.
 
Test results (i.e.\ IP property predictions) are stored in a standard format conforming to schema defined by KIM Property Definitions \cite{kimprops} so that they can be compared within the OpenKIM system with corresponding first-principles and experimental reference data, and can be queried by users through a public facing MongoDB instance. This facilitates analyses across large samples of IPs, such as the study of GB energy in this paper and in \cite{watersAutomatedDeterminationGrain2023}. 


\section{Grain boundary energy simulations}
The grain boundary energy simulations, extracted from OpenKIM, are shown in the following figures. 
See \cite{watersAutomatedDeterminationGrain2023} for model details. 
The average GB energy is shown, along with the lattice-matching (LM) model using the average empirically determined scaling coefficient.
Ag (Figure \ref{fig:Ag}), Al (Figure \ref{fig:Al}), Au (Figure \ref{fig:Au}), Cu (Figure \ref{fig:Cu}), Fe (Figure \ref{fig:Fe}), Ni (Figure \ref{fig:Ni}), Pb (Figure \ref{fig:Pb}), Pd (Figure \ref{fig:Pd}), Pt (Figure \ref{fig:Pt}) and Rh (Figure \ref{fig:Rh}).

\begin{figure}[h]
    \centering
    \includegraphics[width=6in]{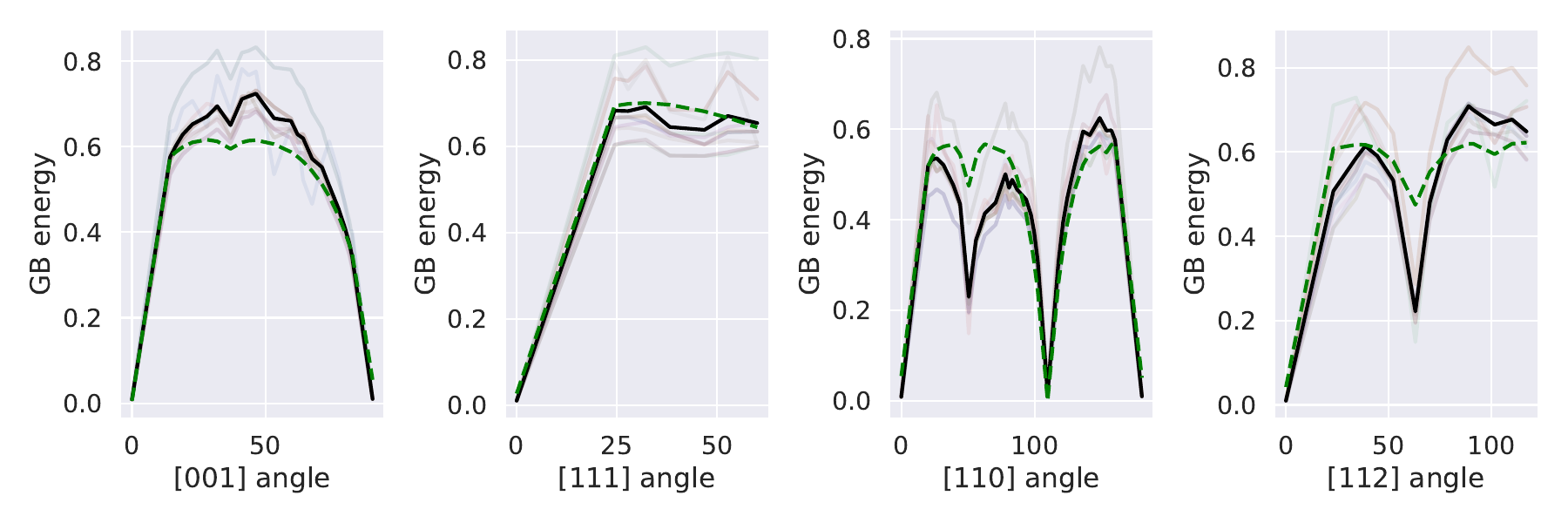}
    \caption{Grain boundary energy simulations for Ag. The average atomistic GB energy from OpenKIM (solid black) is compared with the average best fit analytical LM model (dashed green).}
    \label{fig:Ag}
\end{figure}

\begin{figure}
    \centering
    \includegraphics[width=6in]{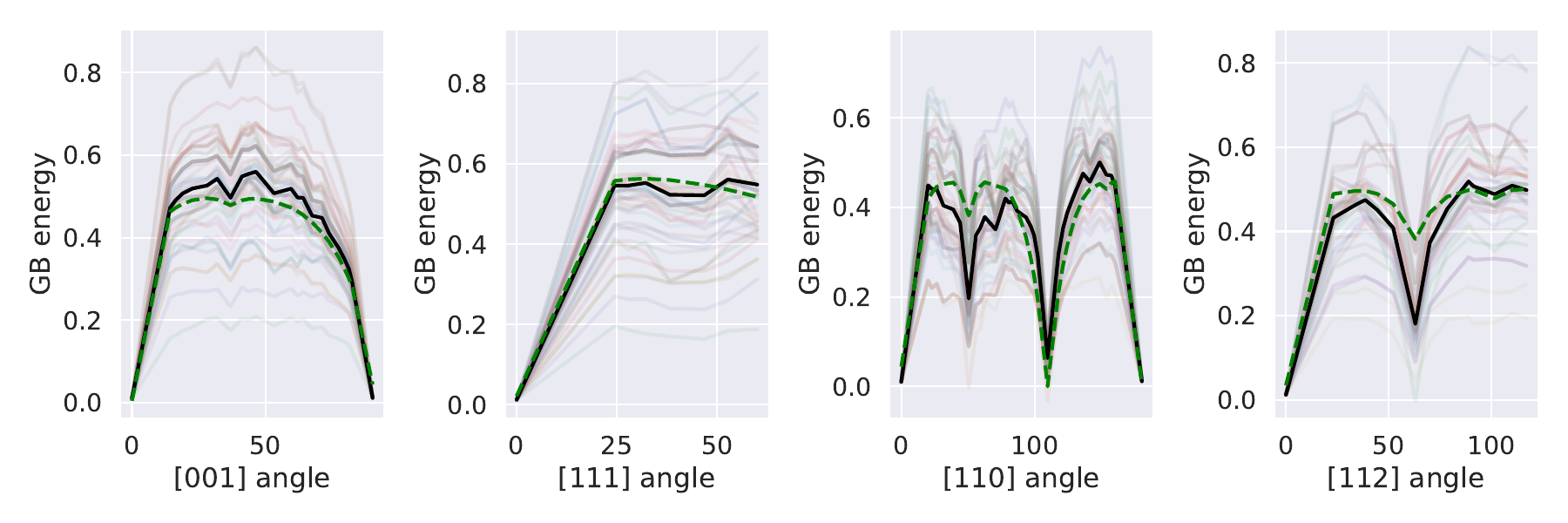}
    \caption{Grain boundary energy simulations for Al. The average atomistic GB energy from OpenKIM (solid black) is compared with the average best fit analytical LM model (dashed green).}
    \label{fig:Al}
\end{figure}

\begin{figure}
    \centering
    \includegraphics[width=6in]{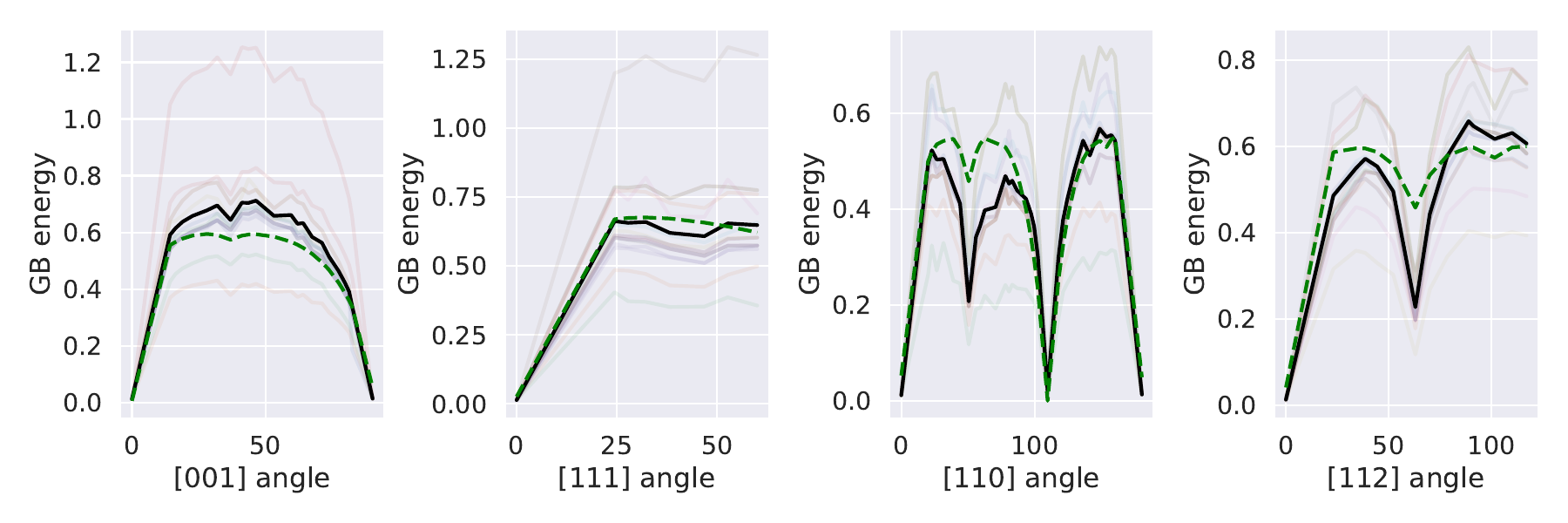}
    \caption{Grain boundary energy simulations for Au. The average atomistic GB energy from OpenKIM (solid black) is compared with the average best fit analytical LM model (dashed green).}
    \label{fig:Au}
\end{figure}

\begin{figure}
    \centering
    \includegraphics[width=6in]{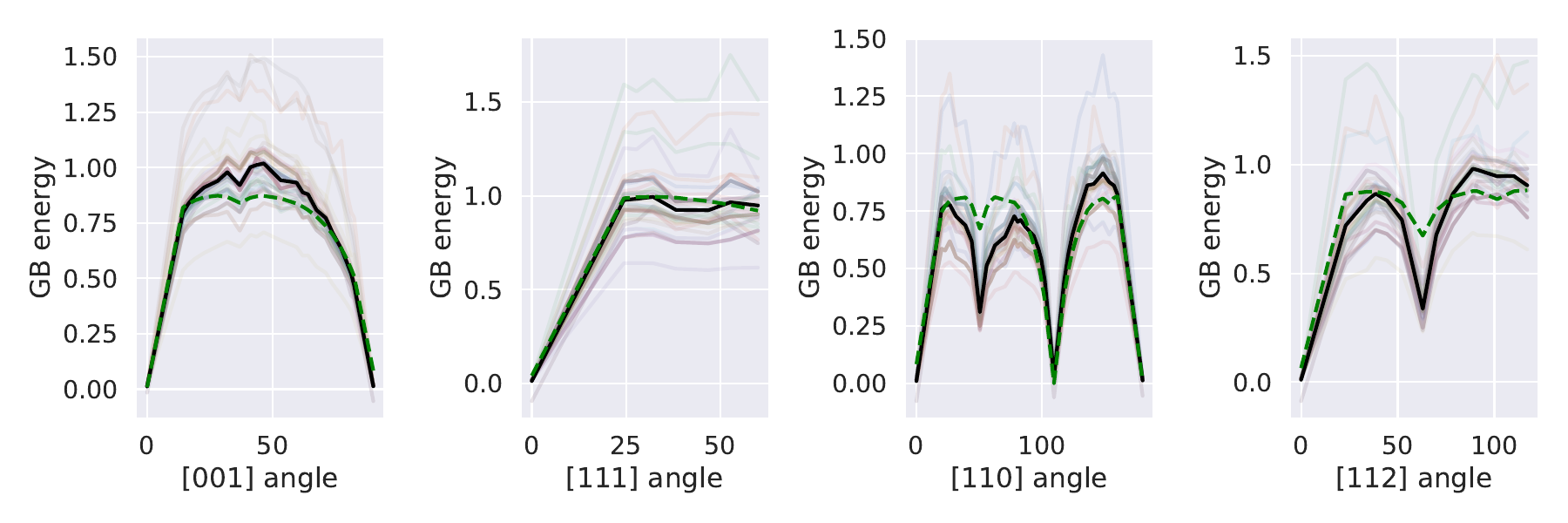}
    \caption{Grain boundary energy simulations for Cu. The average atomistic GB energy from OpenKIM (solid black) is compared with the average best fit analytical LM model (dashed green).}
    \label{fig:Cu}
\end{figure}

\begin{figure}
    \centering
    \includegraphics[width=6in]{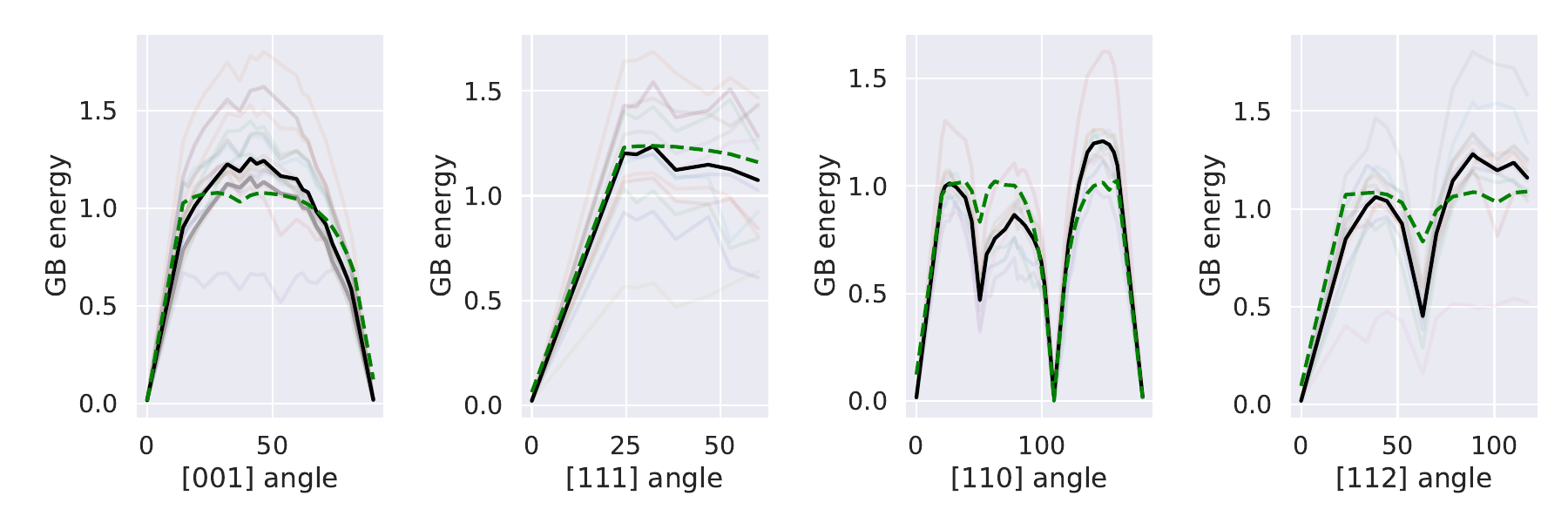}
    \caption{Grain boundary energy simulations for Fe. The average atomistic GB energy from OpenKIM (solid black) is compared with the average best fit analytical LM model (dashed green).}
    \label{fig:Fe}
\end{figure}

\begin{figure}
    \centering
    \includegraphics[width=6in]{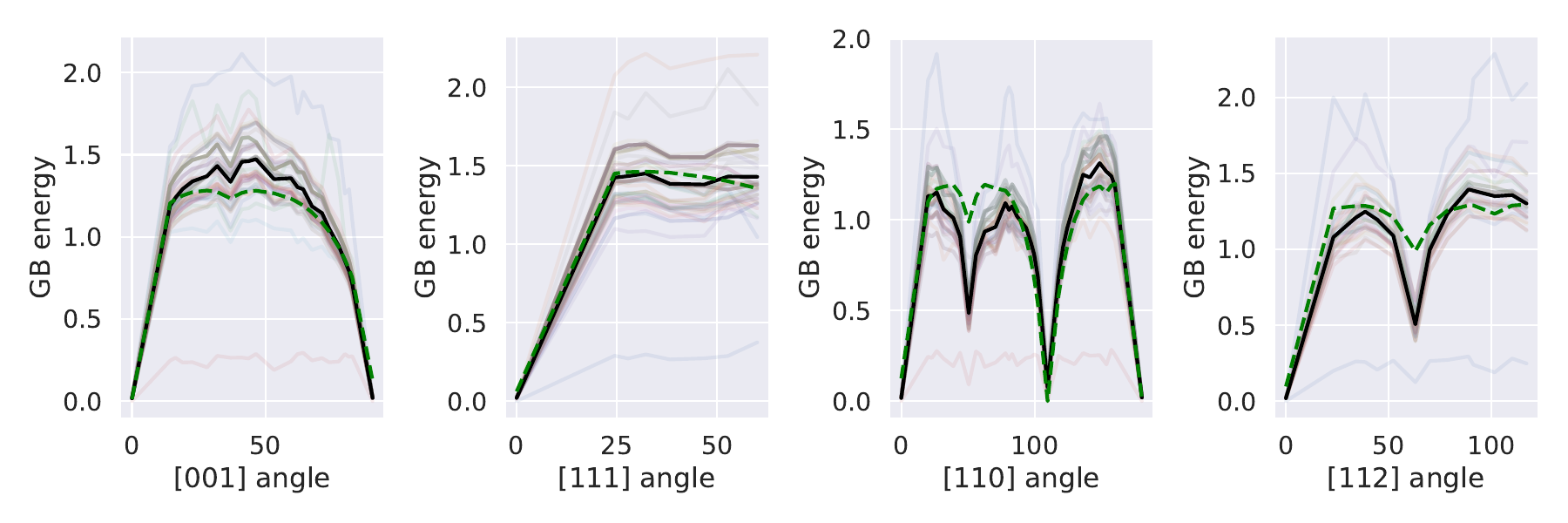}
    \caption{Grain boundary energy simulations for Ni. The average atomistic GB energy from OpenKIM (solid black) is compared with the average best fit analytical LM model (dashed green).}
    \label{fig:Ni}
\end{figure}

\begin{figure}
    \centering
    \includegraphics[width=6in]{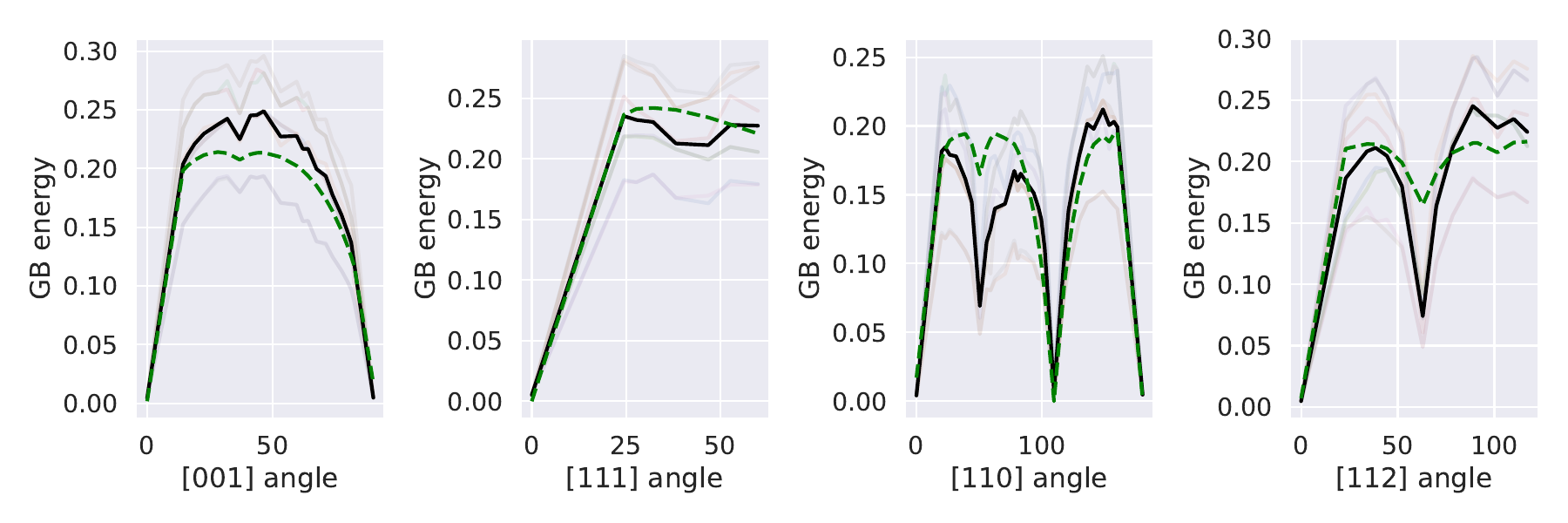}
    \caption{Grain boundary energy simulations for Pb. The average atomistic GB energy from OpenKIM (solid black) is compared with the average best fit analytical LM model (dashed green).}
    \label{fig:Pb}
\end{figure}

\begin{figure}
    \centering
    \includegraphics[width=6in]{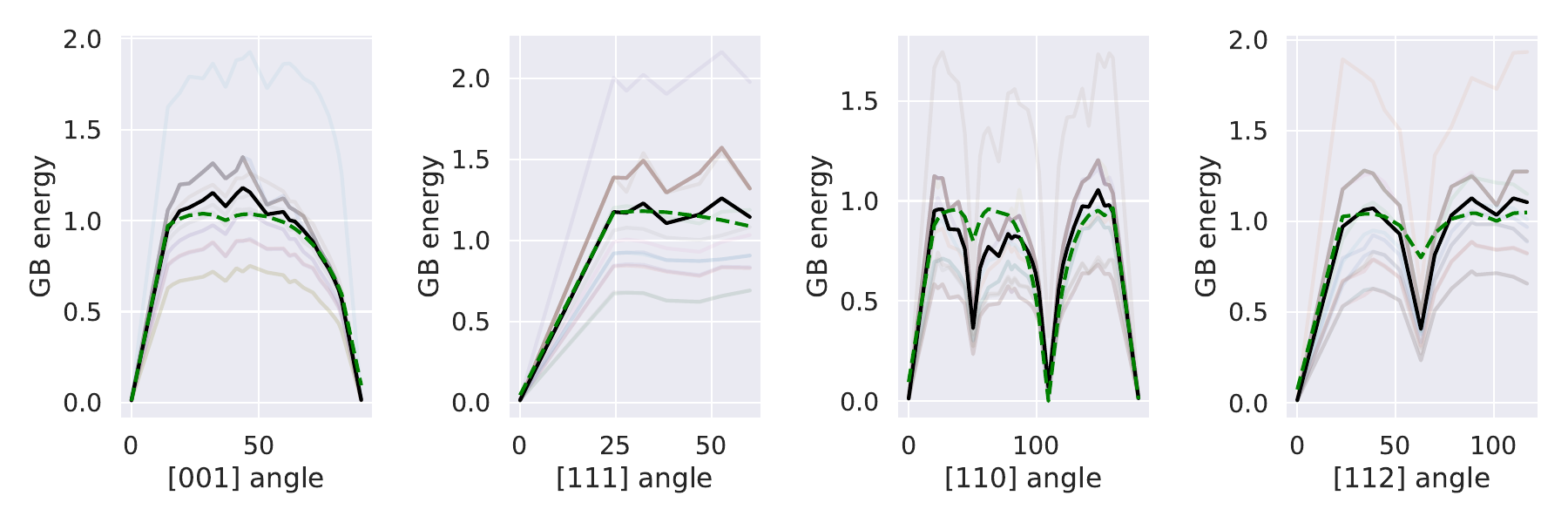}
    \caption{Grain boundary energy simulations for Pd. The average atomistic GB energy from OpenKIM (solid black) is compared with the average best fit analytical LM model (dashed green).}
    \label{fig:Pd}
\end{figure}

\begin{figure}
    \centering
    \includegraphics[width=6in]{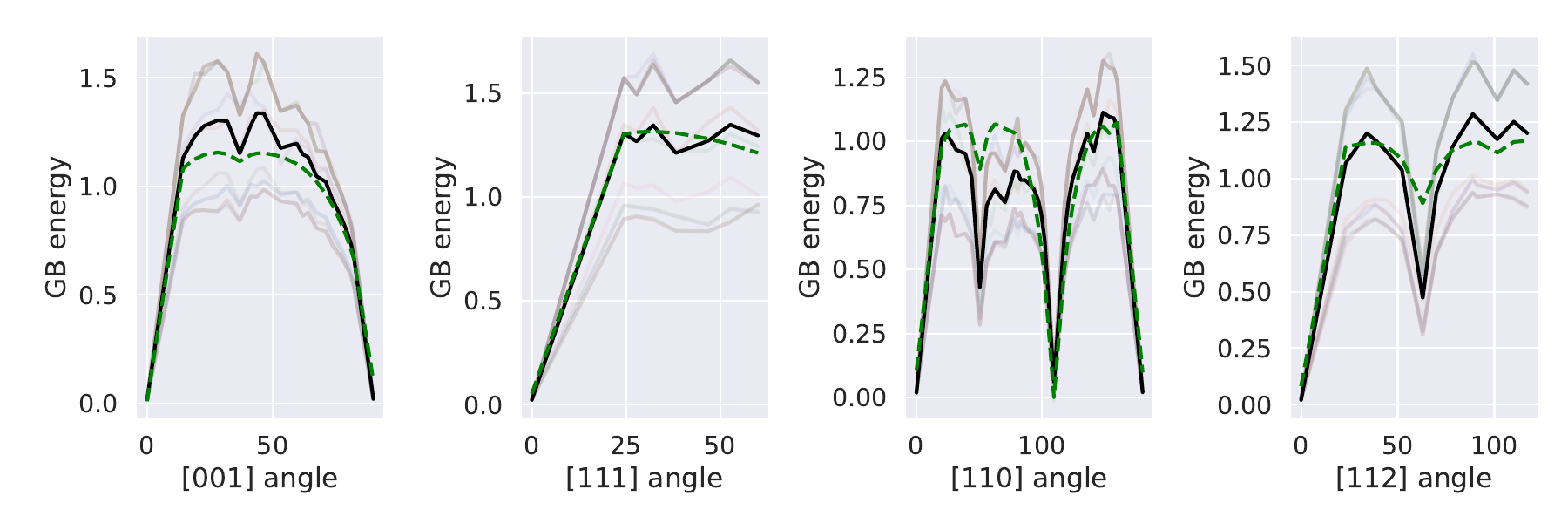}
    \caption{Grain boundary energy simulations for Pt. The average atomistic GB energy from OpenKIM (solid black) is compared with the average best fit analytical LM model (dashed green).}
    \label{fig:Pt}
\end{figure}

\begin{figure}
    \centering
    \includegraphics[width=6in]{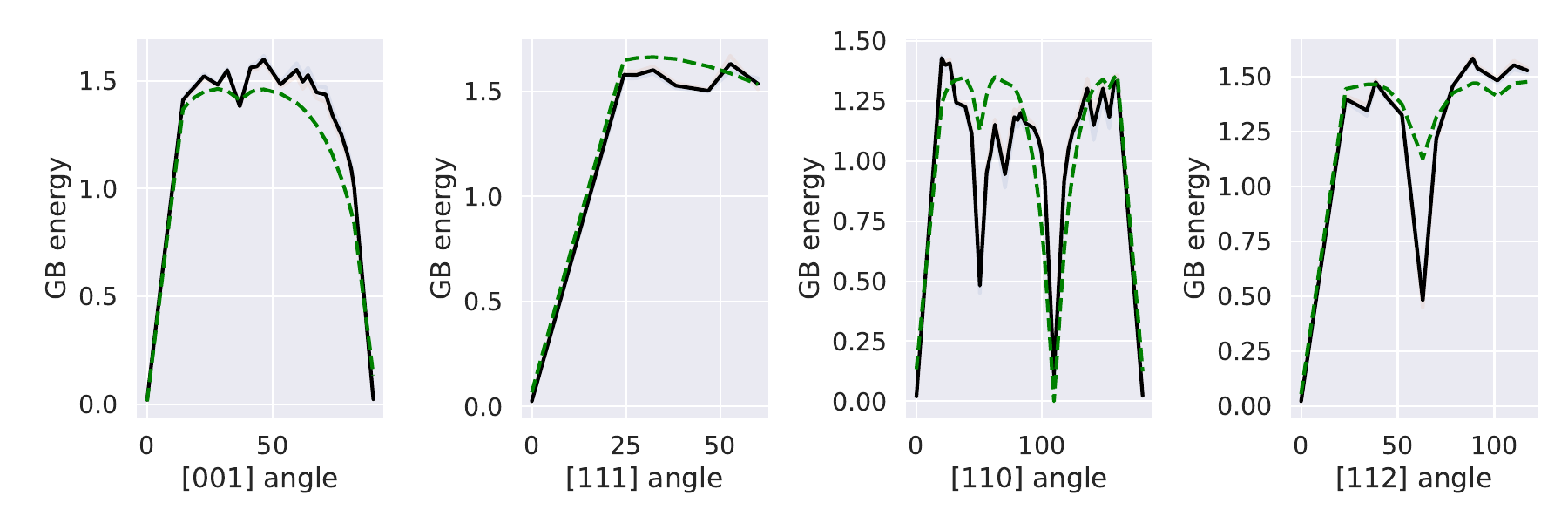}
    \caption{Grain boundary energy simulations for Rh. The average atomistic GB energy from OpenKIM (solid black) is compared with the average best fit analytical LM model (dashed green).}
    \label{fig:Rh}
\end{figure}

\newpage
\section{Property cutoff values for outlier removal}
\label{si:gb:cutoffs}

Property tests are automatically run for relevant models uploaded to OpenKIM.
A property calculation may result in an outlier value for a given model for a number of reasons.
For example, property predictions may be generated automatically by the KIM pipeline for crystal structures that are not intended in the original model fitting.
As a result, it is inevitable that some property calculations will result what can reasonably be considered outlier values. 
The list below outlines the outlier cutoff values used in generating the training dataset for model development.
Boxplots (\Cref{fig:vme_bp,fig:ec_bp,fig:sfe_bp}) of the original property values are provided as justification of the imposed cutoff values.
\begin{enumerate}
	\item All bulk modulus values greater than 100,000 GPa.
	\item All absolute vacancy migration energies greater than 100 eV.
	\item All elastic constants greater than 100,000 GPa.
	\item All stacking fault energies greater than 0.45 ev/A\textsuperscript{2}.
\end{enumerate}

\begin{figure}
    \centering
    \includegraphics{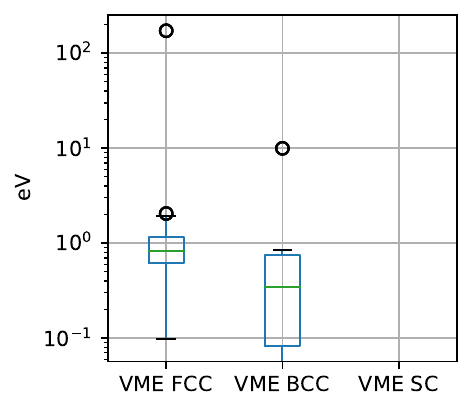}
    \caption{Vacancy migration energy from the original dataset, extracted from OpenKIM for the IPs used. The outlier cutoff of 100 is well above what can be considered a reasonable ``outlier'' limit.}
    \label{fig:vme_bp}
\end{figure}

\begin{figure}
    \centering
    \includegraphics{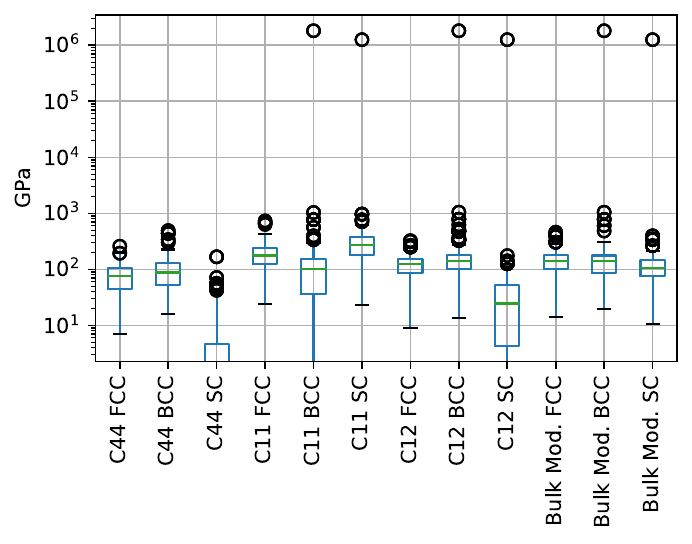}
    \caption{Elastic constants from the original dataset, extracted from OpenKIM for the IPs used. The outlier cutoff of 100,000 is well above what can be considered a reasonable ``outlier'' limit.}
    \label{fig:ec_bp}
\end{figure}

\begin{figure}
    \centering
    \includegraphics{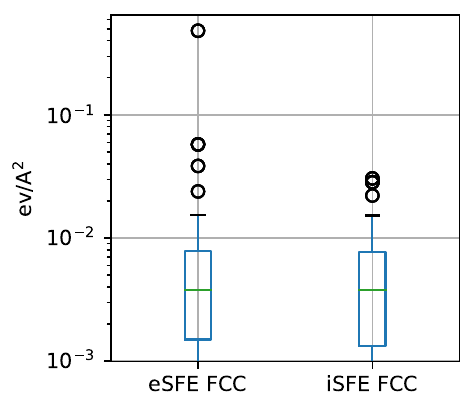}
    \caption{Stacking fault energies from the original dataset, extracted from OpenKIM, for the IPs used. The outlier cutoff of 0.45 is well above what can be considered a reasonable ``outlier'' limit.}
    \label{fig:sfe_bp}
\end{figure}

\newpage
\section{Full correlation heatmap}
The full correlation heatmap is shown in Figure \ref{fig:corr_heatmap_full}.
\begin{figure}
    \centering
    \includegraphics{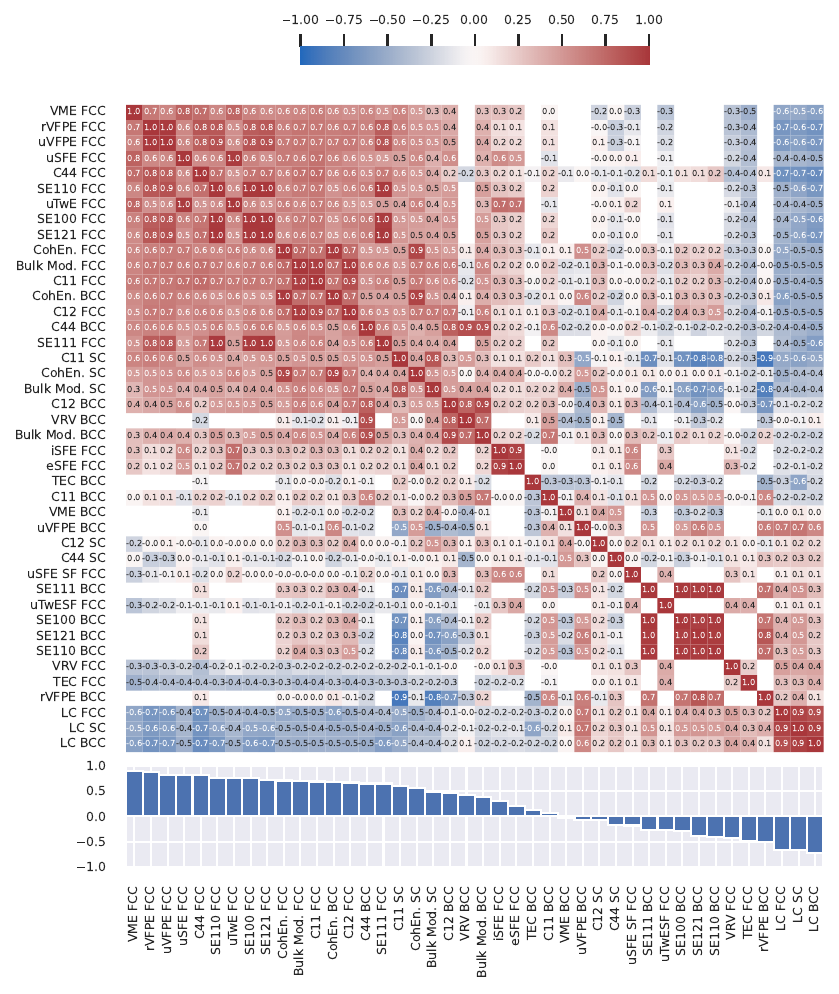}
    \caption{Full correlation heatmap of canonical properties. Blank squares indicate no overlap of calculated values between properties.}
    \label{fig:corr_heatmap_full}
\end{figure}


\newpage
\section{Strongly correlated canonical properties}
Additional supporting pairplots are provided in \Cref{fig:si_gb_pp_bm_elastic,fig:si_gb_pp_se,fig:si_gb_pp_vac,fig:si_gb_lc,fig:si_gb_pp_sfe_twin}.

\begin{figure}[]
	\centering
	\includegraphics[width=4in]{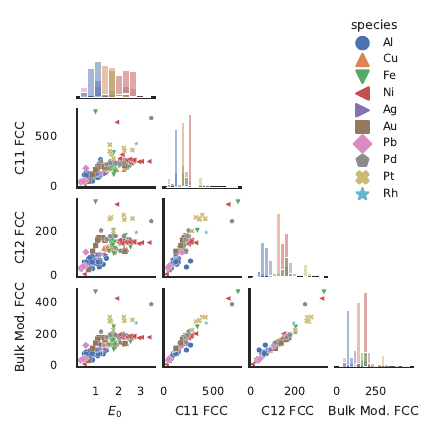}
	\caption{Pairplot of $E_0$ GB scaling coefficient, bulk modulus, and elastic constants. Strong correlation between bulk modulus, $c_{11}$ and $c_{12}$ can be seen.}
	\label{fig:si_gb_pp_bm_elastic}
\end{figure}

\begin{figure}[]
	\centering
	\includegraphics[width=4in]{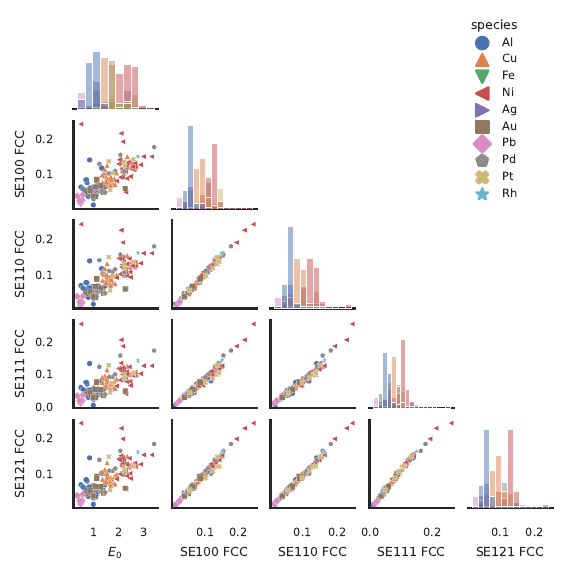}
	\caption{Pairplot of $E_0$ GB scaling coefficient and surface energies. Strong correlation between the surface energies for different tilt axes can be seen.}
	\label{fig:si_gb_pp_se}
\end{figure}

\begin{figure}[]
	\centering
	\includegraphics[width=5in]{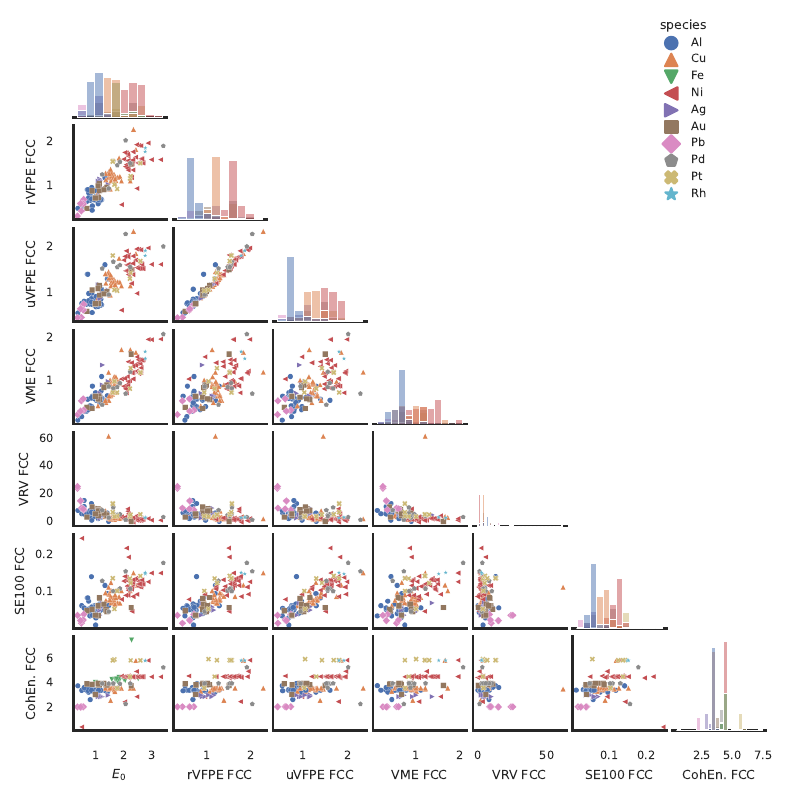}
	\caption{Pairplot of $E_0$ GB scaling coefficient, relaxed and unrelaxed vacancy formation potential energies, vacancy migration energy and vacancy relaxation volume. Strong correlation between rVFPE and uVFPE can be seen.}
	\label{fig:si_gb_pp_vac}
\end{figure}

\begin{figure}[]
	\centering
	\includegraphics[width=4in]{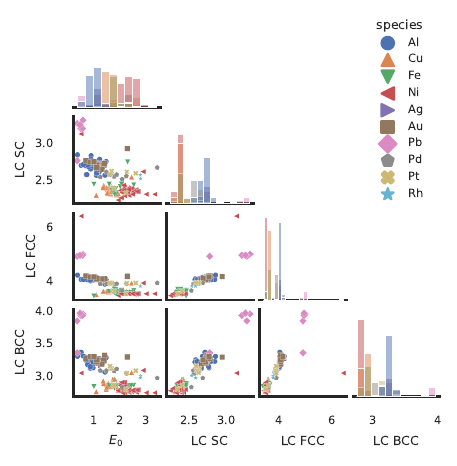}
	\caption{Pairplot of $E_0$ GB scaling coefficient and lattice constants. Strong correlation between LCs for different crystal types can be seen.}
	\label{fig:si_gb_lc}
\end{figure}

\begin{figure}[]
	\centering
	\includegraphics[width=5in]{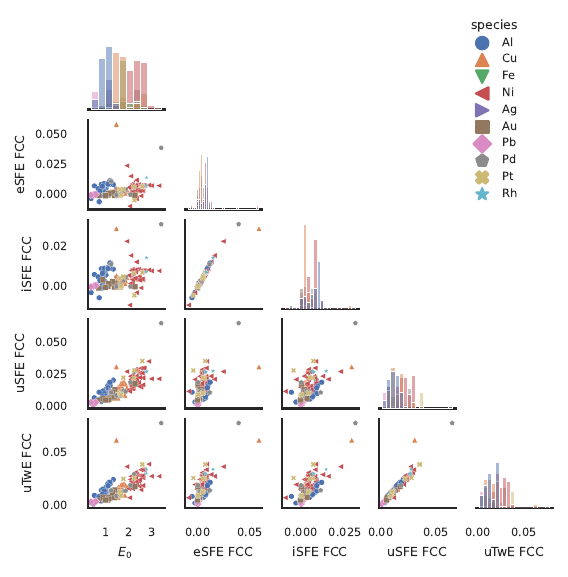}
	\caption{Pairplot of $E_0$ GB scaling coefficient, stacking fault energies and twinning energies. Strong correlation between rVFPE and uVFPE can be seen.}
	\label{fig:si_gb_pp_sfe_twin}
\end{figure}

\newpage
\section{DFT properties}
\Cref{fig:dft_indicator_props} presents a boxplot representation of the canonical property values calculated from IPs versus the DFT calculated values.

\begin{figure}
    \centering
    \subfloat[\centering]{{\includegraphics[width=3in]{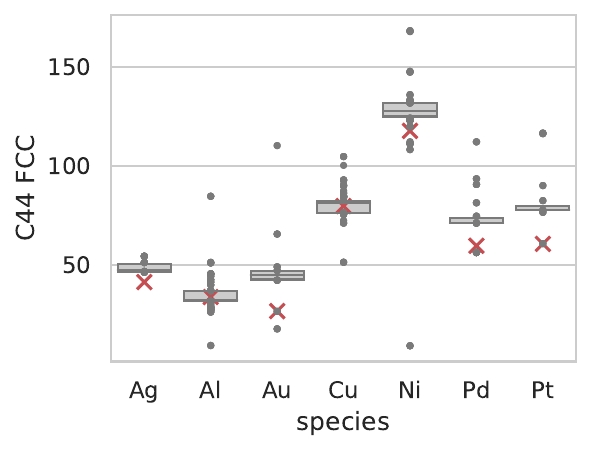}}}
    \subfloat[\centering]{{\includegraphics[width=3in]{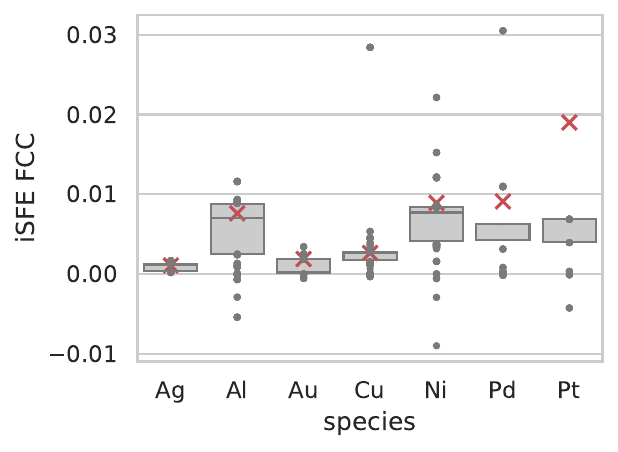}}} \\
    \subfloat[\centering]{{\includegraphics[width=3in]{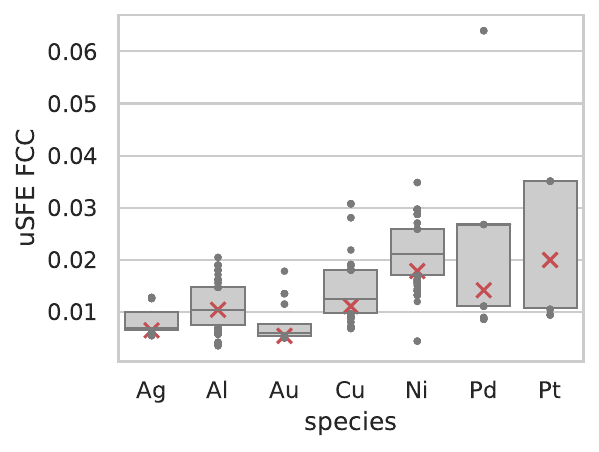}}}
    \caption{
    Comparison of calculated canonical properties. Red X shows the canonical property value calculated using density functional theory (DFT) for three predictors: $C_{44}$ elastic constant, intrinsic stacking fault energy, and unstable stacking fault energy (all FCC). Gray boxes show quartiles of calculated canonical property values from the IP model dataset. Gray circles indicate calculated scaling coefficient results that are outside of the quartiles.
    }
    \label{fig:dft_indicator_props}
\end{figure}

\newpage
\section{Models used}

\begin{table}[]
\centering
\caption{Count of unique models used in this work, broken down by interatomic potential form.}
\label{tab:model_type}
\begin{tabular}{ll}
model type & count \\ \hline
EAM & 108 \\
MEAM & 96 \\
EMT & 10 \\
ADP & 8 \\
Tersoff & 6 \\
SNAP & 4 \\
BOP & 2 \\
ReaxFF & 1
\end{tabular}
\end{table}

The models and drivers used in this work, summarized in \cref{tab:model_type}, were originally published and also archived in OpenKIM through the following references: 
Al \cite{OpenKIM-SM:721930391003:000a, OpenKIM-SM:721930391003:000, OpenKIM-SM:811588957187:000a, OpenKIM-SM:811588957187:000, OpenKIM-SM:871795249052:000a, OpenKIM-SM:871795249052:000, OpenKIM-SM:656517352485:000a, OpenKIM-SM:656517352485:000, OpenKIM-MO:418978237058:005a, OpenKIM-MO:418978237058:005b, OpenKIM-MO:418978237058:005, OpenKIM-MD:120291908751:005, OpenKIM-MO:042691367780:000a, OpenKIM-MO:042691367780:000, OpenKIM-MD:120291908751:005, OpenKIM-MO:699137396381:005a, OpenKIM-MO:699137396381:005b, OpenKIM-MO:699137396381:005c, OpenKIM-MO:699137396381:005, OpenKIM-MD:120291908751:005, OpenKIM-MO:338600200739:000a, OpenKIM-MO:338600200739:000, OpenKIM-MD:120291908751:005, OpenKIM-MO:577453891941:005a, OpenKIM-MO:577453891941:005, OpenKIM-MD:120291908751:005, OpenKIM-MO:658278549784:005a, OpenKIM-MO:658278549784:005, OpenKIM-MD:120291908751:005, OpenKIM-MO:106969701023:005a, OpenKIM-MO:106969701023:005, OpenKIM-MD:120291908751:005, OpenKIM-MO:101214310689:005a, OpenKIM-MO:101214310689:005, OpenKIM-MD:120291908751:005, OpenKIM-MO:751354403791:005a, OpenKIM-MO:751354403791:005, OpenKIM-MD:120291908751:005, OpenKIM-MO:149316865608:005a, OpenKIM-MO:149316865608:005b, OpenKIM-MO:149316865608:005c, OpenKIM-MO:149316865608:005, OpenKIM-MD:120291908751:005, OpenKIM-MO:284963179498:005a, OpenKIM-MO:284963179498:005, OpenKIM-MD:120291908751:005, OpenKIM-MO:122703700223:003a, OpenKIM-MO:122703700223:003, OpenKIM-MD:113599595631:003, OpenKIM-MO:131650261510:005a, OpenKIM-MO:131650261510:005b, OpenKIM-MO:131650261510:005, OpenKIM-MD:120291908751:005, OpenKIM-MO:115316750986:001a, OpenKIM-MO:115316750986:001, OpenKIM-MD:128315414717:004, OpenKIM-MO:623376124862:001a, OpenKIM-MO:623376124862:001, OpenKIM-MD:128315414717:004, OpenKIM-MO:820335782779:000a, OpenKIM-MO:820335782779:000, OpenKIM-MD:120291908751:005, OpenKIM-MO:019873715786:000a, OpenKIM-MO:019873715786:000, OpenKIM-MD:120291908751:005, OpenKIM-MO:020851069572:000a, OpenKIM-MO:020851069572:000, OpenKIM-MD:120291908751:005, OpenKIM-MO:942551040047:005a, OpenKIM-MO:942551040047:005, OpenKIM-MD:120291908751:005, OpenKIM-MO:678952612413:000a, OpenKIM-MO:678952612413:000, OpenKIM-MD:120291908751:005, OpenKIM-MO:651801486679:005a, OpenKIM-MO:651801486679:005, OpenKIM-MD:120291908751:005, OpenKIM-MO:109933561507:005a, OpenKIM-MO:109933561507:005, OpenKIM-MD:120291908751:005, OpenKIM-MO:826591359508:000a, OpenKIM-MO:826591359508:000, OpenKIM-MD:120291908751:005, OpenKIM-MO:324507536345:003a, OpenKIM-MO:324507536345:003, OpenKIM-MO:060567868558:000a, OpenKIM-MO:060567868558:000b, OpenKIM-MO:060567868558:000, OpenKIM-MD:120291908751:005, OpenKIM-MO:722733117926:000a, OpenKIM-MO:722733117926:000, OpenKIM-MD:120291908751:005, OpenKIM-MO:664470114311:005a, OpenKIM-MO:664470114311:005, OpenKIM-MD:120291908751:005, OpenKIM-MO:120808805541:005a, OpenKIM-MO:120808805541:005, OpenKIM-MD:120291908751:005, OpenKIM-MO:878712978062:003a, OpenKIM-MO:878712978062:003, OpenKIM-MD:113599595631:003, OpenKIM-MO:519613893196:000a, OpenKIM-MO:519613893196:000, OpenKIM-MD:120291908751:005, OpenKIM-MO:128037485276:003a, OpenKIM-MO:128037485276:003, OpenKIM-MD:113599595631:003, OpenKIM-MO:117656786760:005a, OpenKIM-MO:117656786760:005, OpenKIM-MD:120291908751:005, OpenKIM-MO:049243498555:000a, OpenKIM-MO:049243498555:000b, OpenKIM-MO:049243498555:000, OpenKIM-MD:120291908751:005, OpenKIM-MO:279544746097:004a, OpenKIM-MO:279544746097:004, OpenKIM-MD:552566534109:004, OpenKIM-MO:992900971352:000a, OpenKIM-MO:992900971352:000, OpenKIM-MD:077075034781:005, OpenKIM-MO:736419017411:000a, OpenKIM-MO:736419017411:000, OpenKIM-MD:077075034781:005, OpenKIM-SM:667696763561:000a, OpenKIM-SM:667696763561:000, OpenKIM-MO:411898953661:004a, OpenKIM-MO:411898953661:004, OpenKIM-MD:552566534109:004, OpenKIM-SM:113843830602:000a, OpenKIM-SM:113843830602:000, OpenKIM-MO:140175748626:004a, OpenKIM-MO:140175748626:004, OpenKIM-MD:552566534109:004, OpenKIM-MO:099716416216:002a, OpenKIM-MO:099716416216:002, OpenKIM-MD:249792265679:002, OpenKIM-MO:478967255435:002a, OpenKIM-MO:478967255435:002, OpenKIM-MD:249792265679:002, OpenKIM-MO:971738391444:001a, OpenKIM-MO:971738391444:001, OpenKIM-MD:249792265679:002, OpenKIM-MO:876687166519:002a, OpenKIM-MO:876687166519:002, OpenKIM-MD:249792265679:002, OpenKIM-MO:616482358807:002a, OpenKIM-MO:616482358807:002, OpenKIM-MD:249792265679:002, OpenKIM-MO:332211522050:002a, OpenKIM-MO:332211522050:002, OpenKIM-MD:249792265679:002, OpenKIM-MO:058537087384:002a, OpenKIM-MO:058537087384:002, OpenKIM-MD:249792265679:002, OpenKIM-MO:618133763375:002a, OpenKIM-MO:618133763375:002, OpenKIM-MD:249792265679:002, OpenKIM-MO:344724145339:002a, OpenKIM-MO:344724145339:002, OpenKIM-MD:249792265679:002, OpenKIM-MO:958395190627:002a, OpenKIM-MO:958395190627:002, OpenKIM-MD:249792265679:002, OpenKIM-MO:596300673917:002a, OpenKIM-MO:596300673917:002, OpenKIM-MD:249792265679:002, OpenKIM-MO:304347095149:001a, OpenKIM-MO:304347095149:001, OpenKIM-MD:249792265679:002, OpenKIM-MO:262519520678:002a, OpenKIM-MO:262519520678:002, OpenKIM-MD:249792265679:002, OpenKIM-MO:461927113651:001a, OpenKIM-MO:461927113651:001, OpenKIM-MD:249792265679:002, OpenKIM-MO:093637366498:002a, OpenKIM-MO:093637366498:002, OpenKIM-MD:249792265679:002, OpenKIM-MO:127847080751:002a, OpenKIM-MO:127847080751:002, OpenKIM-MD:249792265679:002, OpenKIM-MO:353977746962:001a, OpenKIM-MO:353977746962:001, OpenKIM-MD:249792265679:002, OpenKIM-MO:022920256108:002a, OpenKIM-MO:022920256108:002, OpenKIM-MD:249792265679:002, OpenKIM-MO:131642768288:002a, OpenKIM-MO:131642768288:002, OpenKIM-MD:249792265679:002, OpenKIM-MO:315820974149:002a, OpenKIM-MO:315820974149:002, OpenKIM-MD:249792265679:002, OpenKIM-MO:793141037706:002a, OpenKIM-MO:793141037706:002, OpenKIM-MD:249792265679:002, OpenKIM-MO:123629422045:005a, OpenKIM-MO:123629422045:005, OpenKIM-MD:120291908751:005}, 
Cu \cite{OpenKIM-SM:239791545509:000a, OpenKIM-SM:239791545509:000, OpenKIM-SM:316120381362:000a, OpenKIM-SM:316120381362:000, OpenKIM-SM:656517352485:000a, OpenKIM-SM:656517352485:000, OpenKIM-SM:404135993060:000a, OpenKIM-SM:404135993060:000, OpenKIM-SM:566399258279:000a, OpenKIM-SM:566399258279:000, OpenKIM-MO:950828638160:000a, OpenKIM-MO:950828638160:000b, OpenKIM-MO:950828638160:000, OpenKIM-MD:120291908751:005, OpenKIM-MO:128703483589:005a, OpenKIM-MO:128703483589:005, OpenKIM-MD:120291908751:005, OpenKIM-MO:270337113239:005a, OpenKIM-MO:270337113239:005, OpenKIM-MD:120291908751:005, OpenKIM-MO:318213562153:000a, OpenKIM-MO:318213562153:000, OpenKIM-MD:120291908751:005, OpenKIM-MO:127245782811:005a, OpenKIM-MO:127245782811:005b, OpenKIM-MO:127245782811:005, OpenKIM-MD:120291908751:005, OpenKIM-MO:115316750986:001a, OpenKIM-MO:115316750986:001, OpenKIM-MD:128315414717:004, OpenKIM-MO:380822813353:000a, OpenKIM-MO:380822813353:000b, OpenKIM-MO:380822813353:000, OpenKIM-MD:120291908751:005, OpenKIM-MO:173787283511:004a, OpenKIM-MO:173787283511:004, OpenKIM-MD:552566534109:004, OpenKIM-MO:151002396060:004a, OpenKIM-MO:151002396060:004, OpenKIM-MD:552566534109:004, OpenKIM-SM:399364650444:000a, OpenKIM-SM:399364650444:000, OpenKIM-MO:673777079812:004a, OpenKIM-MO:673777079812:004, OpenKIM-MD:552566534109:004, OpenKIM-MO:396616545191:001a, OpenKIM-MO:396616545191:001, OpenKIM-MD:128315414717:004, OpenKIM-SM:667696763561:000a, OpenKIM-SM:667696763561:000, OpenKIM-MO:762798677854:000a, OpenKIM-MO:762798677854:000, OpenKIM-MD:120291908751:005, OpenKIM-MO:592013496703:005a, OpenKIM-MO:592013496703:005, OpenKIM-MD:120291908751:005, OpenKIM-MO:547744193826:000a, OpenKIM-MO:547744193826:000b, OpenKIM-MO:547744193826:000, OpenKIM-MD:120291908751:005, OpenKIM-MO:266134052596:000a, OpenKIM-MO:266134052596:000, OpenKIM-MD:120291908751:005, OpenKIM-MO:469343973171:005a, OpenKIM-MO:469343973171:005, OpenKIM-MD:120291908751:005, OpenKIM-MO:803527979660:000a, OpenKIM-MO:803527979660:000, OpenKIM-MD:120291908751:005, OpenKIM-MO:020851069572:000a, OpenKIM-MO:020851069572:000, OpenKIM-MD:120291908751:005, OpenKIM-MO:097471813275:000a, OpenKIM-MO:097471813275:000, OpenKIM-MD:120291908751:005, OpenKIM-MO:228059236215:001a, OpenKIM-MO:228059236215:001b, OpenKIM-MO:228059236215:001c, OpenKIM-MO:228059236215:001, OpenKIM-MD:128315414717:004, OpenKIM-MO:120596890176:005a, OpenKIM-MO:120596890176:005, OpenKIM-MD:120291908751:005, OpenKIM-MO:759493141826:000a, OpenKIM-MO:759493141826:000b, OpenKIM-MO:759493141826:000, OpenKIM-MD:120291908751:005, OpenKIM-MO:179025990738:005a, OpenKIM-MO:179025990738:005, OpenKIM-MD:120291908751:005, OpenKIM-MO:987541074959:001a, OpenKIM-MO:987541074959:001b, OpenKIM-MO:987541074959:001, OpenKIM-MD:128315414717:004, OpenKIM-MO:426403318662:000a, OpenKIM-MO:426403318662:000, OpenKIM-MD:120291908751:005, OpenKIM-MO:748636486270:005a, OpenKIM-MO:748636486270:005, OpenKIM-MD:120291908751:005, OpenKIM-MO:945691923444:005a, OpenKIM-MO:945691923444:005, OpenKIM-MD:120291908751:005, OpenKIM-MO:609260676108:000a, OpenKIM-MO:609260676108:000, OpenKIM-MD:120291908751:005, OpenKIM-MO:346334655118:005a, OpenKIM-MO:346334655118:005, OpenKIM-MD:120291908751:005, OpenKIM-MO:600021860456:005a, OpenKIM-MO:600021860456:005, OpenKIM-MD:120291908751:005, OpenKIM-MO:942551040047:005a, OpenKIM-MO:942551040047:005, OpenKIM-MD:120291908751:005, OpenKIM-MO:657255834688:000a, OpenKIM-MO:657255834688:000, OpenKIM-MD:120291908751:005, OpenKIM-MO:642748370624:000a, OpenKIM-MO:642748370624:000, OpenKIM-MD:120291908751:005, OpenKIM-MO:529419924683:000a, OpenKIM-MO:529419924683:000, OpenKIM-MD:536750310735:000, OpenKIM-MO:227887284491:002a, OpenKIM-MO:227887284491:002, OpenKIM-MD:249792265679:002, OpenKIM-MO:063626065437:002a, OpenKIM-MO:063626065437:002, OpenKIM-MD:249792265679:002, OpenKIM-MO:409065472403:002a, OpenKIM-MO:409065472403:002, OpenKIM-MD:249792265679:002, OpenKIM-MO:849011491644:002a, OpenKIM-MO:849011491644:002, OpenKIM-MD:249792265679:002, OpenKIM-MO:813575892799:002a, OpenKIM-MO:813575892799:002, OpenKIM-MD:249792265679:002, OpenKIM-MO:070797404269:002a, OpenKIM-MO:070797404269:002, OpenKIM-MD:249792265679:002, OpenKIM-MO:353393547686:002a, OpenKIM-MO:353393547686:002, OpenKIM-MD:249792265679:002, OpenKIM-MO:028979335952:002a, OpenKIM-MO:028979335952:002, OpenKIM-MD:249792265679:002, OpenKIM-MO:262519520678:002a, OpenKIM-MO:262519520678:002, OpenKIM-MD:249792265679:002, OpenKIM-MO:694335101831:002a, OpenKIM-MO:694335101831:002, OpenKIM-MD:249792265679:002, OpenKIM-MO:486450342170:002a, OpenKIM-MO:486450342170:002, OpenKIM-MD:249792265679:002, OpenKIM-MO:390178379548:002a, OpenKIM-MO:390178379548:002, OpenKIM-MD:249792265679:002, OpenKIM-MO:407917731909:001a, OpenKIM-MO:407917731909:001, OpenKIM-MD:249792265679:002, OpenKIM-MO:087820130586:001a, OpenKIM-MO:087820130586:001, OpenKIM-MD:249792265679:002, OpenKIM-MO:122936827583:002a, OpenKIM-MO:122936827583:002, OpenKIM-MD:249792265679:002, OpenKIM-MO:380272712420:002a, OpenKIM-MO:380272712420:002, OpenKIM-MD:249792265679:002, OpenKIM-MO:265210066873:000a, OpenKIM-MO:265210066873:000, OpenKIM-MD:536750310735:000}, 
Fe \cite{OpenKIM-SM:267016608755:000a, OpenKIM-SM:267016608755:000, OpenKIM-SM:656517352485:000a, OpenKIM-SM:656517352485:000, OpenKIM-MO:142799717516:005a, OpenKIM-MO:142799717516:005b, OpenKIM-MO:142799717516:005, OpenKIM-MD:120291908751:005, OpenKIM-MO:763197941039:000a, OpenKIM-MO:763197941039:000, OpenKIM-MD:120291908751:005, OpenKIM-MO:044341472608:000a, OpenKIM-MO:044341472608:000, OpenKIM-MD:120291908751:005, OpenKIM-MO:147603128437:004a, OpenKIM-MO:147603128437:004, OpenKIM-MD:552566534109:004, OpenKIM-MO:331285495617:004a, OpenKIM-MO:331285495617:004, OpenKIM-MD:552566534109:004, OpenKIM-MO:984358344196:004a, OpenKIM-MO:984358344196:004, OpenKIM-MD:552566534109:004, OpenKIM-MO:715003088863:000a, OpenKIM-MO:715003088863:000, OpenKIM-MD:120291908751:005, OpenKIM-MO:657255834688:000a, OpenKIM-MO:657255834688:000, OpenKIM-MD:120291908751:005, OpenKIM-MO:820335782779:000a, OpenKIM-MO:820335782779:000, OpenKIM-MD:120291908751:005, OpenKIM-SM:473463498269:000a, OpenKIM-SM:473463498269:000, OpenKIM-MO:857282754307:003a, OpenKIM-MO:857282754307:003, OpenKIM-MO:137964310702:004a, OpenKIM-MO:137964310702:004, OpenKIM-MD:077075034781:005, OpenKIM-MO:608695023236:000a, OpenKIM-MO:608695023236:000, OpenKIM-MD:077075034781:005, OpenKIM-MO:036303866285:000a, OpenKIM-MO:036303866285:000, OpenKIM-MD:120291908751:005, OpenKIM-MO:024705128470:000a, OpenKIM-MO:024705128470:000, OpenKIM-MD:120291908751:005, OpenKIM-MO:135034229282:002a, OpenKIM-MO:135034229282:002b, OpenKIM-MO:135034229282:002, OpenKIM-MD:620624592962:002, OpenKIM-MO:803527979660:000a, OpenKIM-MO:803527979660:000, OpenKIM-MD:120291908751:005, OpenKIM-MO:492310898779:002a, OpenKIM-MO:492310898779:002, OpenKIM-MD:249792265679:002, OpenKIM-MO:549900287421:002a, OpenKIM-MO:549900287421:002, OpenKIM-MD:249792265679:002, OpenKIM-MO:150993986463:001a, OpenKIM-MO:150993986463:001, OpenKIM-MD:249792265679:002, OpenKIM-MO:115454747503:002a, OpenKIM-MO:115454747503:002, OpenKIM-MD:249792265679:002, OpenKIM-MO:924736622203:002a, OpenKIM-MO:924736622203:002, OpenKIM-MD:249792265679:002, OpenKIM-MO:072689718616:002a, OpenKIM-MO:072689718616:002, OpenKIM-MD:249792265679:002, OpenKIM-MO:179420363944:002a, OpenKIM-MO:179420363944:002, OpenKIM-MD:249792265679:002, OpenKIM-MO:332211522050:002a, OpenKIM-MO:332211522050:002, OpenKIM-MD:249792265679:002, OpenKIM-MO:343168101490:002a, OpenKIM-MO:343168101490:002, OpenKIM-MD:249792265679:002, OpenKIM-MO:912636107108:002a, OpenKIM-MO:912636107108:002, OpenKIM-MD:249792265679:002, OpenKIM-MO:063626065437:002a, OpenKIM-MO:063626065437:002, OpenKIM-MD:249792265679:002, OpenKIM-MO:162036141261:002a, OpenKIM-MO:162036141261:002, OpenKIM-MD:249792265679:002, OpenKIM-MO:304347095149:001a, OpenKIM-MO:304347095149:001, OpenKIM-MD:249792265679:002, OpenKIM-MO:321233176498:002a, OpenKIM-MO:321233176498:002, OpenKIM-MD:249792265679:002, OpenKIM-MO:110119204723:002a, OpenKIM-MO:110119204723:002, OpenKIM-MD:249792265679:002, OpenKIM-MO:260546967793:002a, OpenKIM-MO:260546967793:002, OpenKIM-MD:249792265679:002, OpenKIM-MO:262519520678:002a, OpenKIM-MO:262519520678:002, OpenKIM-MD:249792265679:002}, 
Ni \cite{OpenKIM-SM:078420412697:000a, OpenKIM-SM:078420412697:000, OpenKIM-SM:333792531460:000a, OpenKIM-SM:333792531460:000, OpenKIM-SM:770142935022:000a, OpenKIM-SM:770142935022:000, OpenKIM-SM:971529344487:000a, OpenKIM-SM:971529344487:000, OpenKIM-MO:803527979660:000a, OpenKIM-MO:803527979660:000, OpenKIM-MD:120291908751:005, OpenKIM-MO:222110751402:000a, OpenKIM-MO:222110751402:000, OpenKIM-MD:120291908751:005, OpenKIM-MO:400591584784:005a, OpenKIM-MO:400591584784:005, OpenKIM-MD:120291908751:005, OpenKIM-MO:751354403791:005a, OpenKIM-MO:751354403791:005, OpenKIM-MD:120291908751:005, OpenKIM-MO:535504325462:003, OpenKIM-MO:535504325462:003, OpenKIM-MD:120291908751:005, OpenKIM-MO:036303866285:000a, OpenKIM-MO:036303866285:000, OpenKIM-MD:120291908751:005, OpenKIM-MO:103383163946:000a, OpenKIM-MO:103383163946:000, OpenKIM-MD:120291908751:005, OpenKIM-MO:826591359508:000a, OpenKIM-MO:826591359508:000, OpenKIM-MD:120291908751:005, OpenKIM-MO:110256178378:005a, OpenKIM-MO:110256178378:005b, OpenKIM-MO:110256178378:005, OpenKIM-MD:120291908751:005, OpenKIM-MO:115316750986:001a, OpenKIM-MO:115316750986:001, OpenKIM-MD:128315414717:004, OpenKIM-MO:758825945924:004a, OpenKIM-MO:758825945924:004, OpenKIM-MD:552566534109:004, OpenKIM-SM:559286646876:000a, OpenKIM-SM:559286646876:000, OpenKIM-SM:477692857359:000a, OpenKIM-SM:477692857359:000, OpenKIM-MO:381861218831:004a, OpenKIM-MO:381861218831:004, OpenKIM-MD:552566534109:004, OpenKIM-MO:322509103239:004a, OpenKIM-MO:322509103239:004, OpenKIM-MD:552566534109:004, OpenKIM-SM:306597220004:000a, OpenKIM-SM:306597220004:000, OpenKIM-MO:109933561507:005a, OpenKIM-MO:109933561507:005, OpenKIM-MD:120291908751:005, OpenKIM-MO:592013496703:005a, OpenKIM-MO:592013496703:005, OpenKIM-MD:120291908751:005, OpenKIM-MO:101214310689:005a, OpenKIM-MO:101214310689:005, OpenKIM-MD:120291908751:005, OpenKIM-MO:010613863288:000a, OpenKIM-MO:010613863288:000, OpenKIM-MD:120291908751:005, OpenKIM-MO:128037485276:003a, OpenKIM-MO:128037485276:003, OpenKIM-MD:113599595631:003, OpenKIM-MO:532072268679:000a, OpenKIM-MO:532072268679:000, OpenKIM-MD:120291908751:005, OpenKIM-MO:047308317761:000a, OpenKIM-MO:047308317761:000, OpenKIM-MD:120291908751:005, OpenKIM-MO:593762436933:000a, OpenKIM-MO:593762436933:000b, OpenKIM-MO:593762436933:000, OpenKIM-MD:120291908751:005, OpenKIM-MO:306032198193:000a, OpenKIM-MO:306032198193:000, OpenKIM-MD:120291908751:005, OpenKIM-MO:108408461881:001a, OpenKIM-MO:108408461881:001, OpenKIM-MD:128315414717:004, OpenKIM-MO:800536961967:003a, OpenKIM-MO:800536961967:003, OpenKIM-MD:853402641673:002, OpenKIM-MO:769632475533:000a, OpenKIM-MO:769632475533:000, OpenKIM-MD:120291908751:005, OpenKIM-MO:418978237058:005a, OpenKIM-MO:418978237058:005b, OpenKIM-MO:418978237058:005, OpenKIM-MD:120291908751:005, OpenKIM-MO:977363131043:005a, OpenKIM-MO:977363131043:005, OpenKIM-MD:120291908751:005, OpenKIM-MO:267721408934:005a, OpenKIM-MO:267721408934:005, OpenKIM-MD:120291908751:005, OpenKIM-MO:820335782779:000a, OpenKIM-MO:820335782779:000, OpenKIM-MD:120291908751:005, OpenKIM-MO:763197941039:000a, OpenKIM-MO:763197941039:000, OpenKIM-MD:120291908751:005, OpenKIM-MO:266134052596:000a, OpenKIM-MO:266134052596:000, OpenKIM-MD:120291908751:005, OpenKIM-MO:149104665840:005a, OpenKIM-MO:149104665840:005, OpenKIM-MD:120291908751:005, OpenKIM-MO:832600236922:005a, OpenKIM-MO:832600236922:005, OpenKIM-MD:120291908751:005, OpenKIM-MO:871937946490:000a, OpenKIM-MO:871937946490:000, OpenKIM-MD:120291908751:005, OpenKIM-MO:657255834688:000a, OpenKIM-MO:657255834688:000, OpenKIM-MD:120291908751:005, OpenKIM-MO:469343973171:005a, OpenKIM-MO:469343973171:005, OpenKIM-MD:120291908751:005, OpenKIM-MO:715003088863:000a, OpenKIM-MO:715003088863:000, OpenKIM-MD:120291908751:005, OpenKIM-MO:776437554506:000a, OpenKIM-MO:776437554506:000, OpenKIM-MD:120291908751:005, OpenKIM-MO:937008984446:002a, OpenKIM-MO:937008984446:002, OpenKIM-MD:249792265679:002, OpenKIM-MO:876687166519:002a, OpenKIM-MO:876687166519:002, OpenKIM-MD:249792265679:002, OpenKIM-MO:700541006254:002a, OpenKIM-MO:700541006254:002, OpenKIM-MD:249792265679:002, OpenKIM-MO:091278480940:002a, OpenKIM-MO:091278480940:002, OpenKIM-MD:249792265679:002, OpenKIM-MO:115454747503:002a, OpenKIM-MO:115454747503:002, OpenKIM-MD:249792265679:002, OpenKIM-MO:663355627503:002a, OpenKIM-MO:663355627503:002, OpenKIM-MD:249792265679:002, OpenKIM-MO:500937681860:002a, OpenKIM-MO:500937681860:002, OpenKIM-MD:249792265679:002, OpenKIM-MO:880803040302:002a, OpenKIM-MO:880803040302:002, OpenKIM-MD:249792265679:002, OpenKIM-MO:020840179467:002a, OpenKIM-MO:020840179467:002, OpenKIM-MD:249792265679:002, OpenKIM-MO:131642768288:002a, OpenKIM-MO:131642768288:002, OpenKIM-MD:249792265679:002, OpenKIM-MO:461927113651:001a, OpenKIM-MO:461927113651:001, OpenKIM-MD:249792265679:002, OpenKIM-MO:744610363128:002a, OpenKIM-MO:744610363128:002, OpenKIM-MD:249792265679:002, OpenKIM-MO:478967255435:002a, OpenKIM-MO:478967255435:002, OpenKIM-MD:249792265679:002, OpenKIM-MO:612225165948:002a, OpenKIM-MO:612225165948:002, OpenKIM-MD:249792265679:002, OpenKIM-MO:000553624872:001a, OpenKIM-MO:000553624872:001, OpenKIM-MD:249792265679:002, OpenKIM-MO:050461957184:002a, OpenKIM-MO:050461957184:002, OpenKIM-MD:249792265679:002, OpenKIM-MO:348689608050:002a, OpenKIM-MO:348689608050:002, OpenKIM-MD:249792265679:002, OpenKIM-MO:912636107108:002a, OpenKIM-MO:912636107108:002, OpenKIM-MD:249792265679:002, OpenKIM-MO:409065472403:002a, OpenKIM-MO:409065472403:002, OpenKIM-MD:249792265679:002, OpenKIM-MO:321233176498:002a, OpenKIM-MO:321233176498:002, OpenKIM-MD:249792265679:002, OpenKIM-MO:913991514986:000a, OpenKIM-MO:913991514986:000, OpenKIM-MD:536750310735:000, OpenKIM-MO:468686727341:000a, OpenKIM-MO:468686727341:000, OpenKIM-MD:536750310735:000, OpenKIM-MO:008996216289:002a, OpenKIM-MO:008996216289:002, OpenKIM-MD:249792265679:002, OpenKIM-MO:000553624872:000a, OpenKIM-MO:000553624872:000, OpenKIM-MD:249792265679:001}, 
Ag \cite{OpenKIM-MO:137719994600:004a, OpenKIM-MO:137719994600:004, OpenKIM-MD:552566534109:004, OpenKIM-MO:112077942578:002a, OpenKIM-MO:112077942578:002, OpenKIM-MD:249792265679:002, OpenKIM-MO:115316750986:001a, OpenKIM-MO:115316750986:001, OpenKIM-MD:128315414717:004, OpenKIM-MO:861893969202:004a, OpenKIM-MO:861893969202:004, OpenKIM-MD:552566534109:004, OpenKIM-MO:111986436268:004a, OpenKIM-MO:111986436268:004, OpenKIM-MD:552566534109:004, OpenKIM-MO:969318541747:001a, OpenKIM-MO:969318541747:001, OpenKIM-MD:249792265679:002, OpenKIM-MO:813575892799:002a, OpenKIM-MO:813575892799:002, OpenKIM-MD:249792265679:002, OpenKIM-MO:303974873468:001a, OpenKIM-MO:303974873468:001, OpenKIM-MD:128315414717:004, OpenKIM-MO:222110751402:000a, OpenKIM-MO:222110751402:000, OpenKIM-MD:120291908751:005, OpenKIM-MO:505250810900:000a, OpenKIM-MO:505250810900:000b, OpenKIM-MO:505250810900:000, OpenKIM-MD:120291908751:005, OpenKIM-MO:108983864770:005a, OpenKIM-MO:108983864770:005b, OpenKIM-MO:108983864770:005, OpenKIM-MD:120291908751:005, OpenKIM-MO:318213562153:000a, OpenKIM-MO:318213562153:000, OpenKIM-MD:120291908751:005, OpenKIM-MO:270337113239:005a, OpenKIM-MO:270337113239:005, OpenKIM-MD:120291908751:005, OpenKIM-MO:947112899505:005a, OpenKIM-MO:947112899505:005b, OpenKIM-MO:947112899505:005, OpenKIM-MD:120291908751:005, OpenKIM-MO:104806802344:005a, OpenKIM-MO:104806802344:005b, OpenKIM-MO:104806802344:005, OpenKIM-MD:120291908751:005, OpenKIM-MO:212700056563:005a, OpenKIM-MO:212700056563:005, OpenKIM-MD:120291908751:005, OpenKIM-MO:131620013077:005a, OpenKIM-MO:131620013077:005, OpenKIM-MD:120291908751:005, OpenKIM-MO:055919219575:000a, OpenKIM-MO:055919219575:000, OpenKIM-MD:120291908751:005, OpenKIM-MO:128703483589:005a, OpenKIM-MO:128703483589:005, OpenKIM-MD:120291908751:005}, 
Au \cite{OpenKIM-MO:684444719999:000a, OpenKIM-MO:684444719999:000b, OpenKIM-MO:684444719999:000, OpenKIM-MD:120291908751:005, OpenKIM-MO:104891429740:005a, OpenKIM-MO:104891429740:005b, OpenKIM-MO:104891429740:005c, OpenKIM-MO:104891429740:005, OpenKIM-MD:120291908751:005, OpenKIM-MO:188701096956:000a, OpenKIM-MO:188701096956:000, OpenKIM-MD:120291908751:005, OpenKIM-MO:173248269481:000a, OpenKIM-MO:173248269481:000, OpenKIM-MD:120291908751:005, OpenKIM-MO:468407568810:005a, OpenKIM-MO:468407568810:005b, OpenKIM-MO:468407568810:005, OpenKIM-MD:120291908751:005, OpenKIM-MO:318213562153:000a, OpenKIM-MO:318213562153:000, OpenKIM-MD:120291908751:005, OpenKIM-MO:115316750986:001a, OpenKIM-MO:115316750986:001, OpenKIM-MD:128315414717:004, OpenKIM-MO:754413982908:000a, OpenKIM-MO:754413982908:000, OpenKIM-MD:120291908751:005, OpenKIM-MO:774911580446:001a, OpenKIM-MO:774911580446:001, OpenKIM-MD:249792265679:002, OpenKIM-MO:017524376569:001a, OpenKIM-MO:017524376569:001, OpenKIM-MD:128315414717:004, OpenKIM-SM:066295357485:000a, OpenKIM-SM:066295357485:000, OpenKIM-SM:985135773293:000a, OpenKIM-SM:985135773293:000, OpenKIM-SM:113843830602:000a, OpenKIM-SM:113843830602:000, OpenKIM-SM:974345878378:001a, OpenKIM-SM:974345878378:001, OpenKIM-MO:463728687265:000a, OpenKIM-MO:463728687265:000, OpenKIM-MD:532469991695:003, OpenKIM-MO:946831081299:000a, OpenKIM-MO:946831081299:000, OpenKIM-MD:120291908751:005}, 
Pb \cite{OpenKIM-MO:961101070310:000a, OpenKIM-MO:961101070310:000, OpenKIM-MD:120291908751:005, OpenKIM-MO:699137396381:005a, OpenKIM-MO:699137396381:005b, OpenKIM-MO:699137396381:005c, OpenKIM-MO:699137396381:005, OpenKIM-MD:120291908751:005, OpenKIM-MO:325675357262:002a, OpenKIM-MO:325675357262:002, OpenKIM-MD:249792265679:002, OpenKIM-MO:534638645497:004a, OpenKIM-MO:534638645497:004, OpenKIM-MD:552566534109:004, OpenKIM-MO:988703794028:000a, OpenKIM-MO:988703794028:000b, OpenKIM-MO:988703794028:000, OpenKIM-MD:120291908751:005, OpenKIM-MO:162736908871:002a, OpenKIM-MO:162736908871:002, OpenKIM-MD:249792265679:002, OpenKIM-MO:370271093517:004a, OpenKIM-MO:370271093517:004, OpenKIM-MD:552566534109:004, OpenKIM-MO:116920074573:005a, OpenKIM-MO:116920074573:005b, OpenKIM-MO:116920074573:005, OpenKIM-MD:120291908751:005, OpenKIM-MO:958424213898:004a, OpenKIM-MO:958424213898:004, OpenKIM-MD:552566534109:004, OpenKIM-MO:019208265157:001a, OpenKIM-MO:019208265157:001, OpenKIM-MD:249792265679:002, OpenKIM-MO:119135752160:005a, OpenKIM-MO:119135752160:005, OpenKIM-MD:120291908751:005}, 
Pd \cite{OpenKIM-MO:924736622203:002a, OpenKIM-MO:924736622203:002, OpenKIM-MD:249792265679:002, OpenKIM-MO:068985622065:002a, OpenKIM-MO:068985622065:002, OpenKIM-MD:249792265679:002, OpenKIM-MO:865505436319:000a, OpenKIM-MO:865505436319:000, OpenKIM-MD:120291908751:005, OpenKIM-MO:115316750986:001a, OpenKIM-MO:115316750986:001, OpenKIM-MD:128315414717:004, OpenKIM-MO:066802556726:001a, OpenKIM-MO:066802556726:001, OpenKIM-MD:128315414717:004, OpenKIM-MO:046547823135:002a, OpenKIM-MO:046547823135:002, OpenKIM-MD:249792265679:002, OpenKIM-MO:532072268679:000a, OpenKIM-MO:532072268679:000, OpenKIM-MD:120291908751:005, OpenKIM-SM:559286646876:000a, OpenKIM-SM:559286646876:000, OpenKIM-MO:008996216289:002a, OpenKIM-MO:008996216289:002, OpenKIM-MD:249792265679:002, OpenKIM-MO:993644691224:000a, OpenKIM-MO:993644691224:000b, OpenKIM-MO:993644691224:000, OpenKIM-MD:120291908751:005, OpenKIM-MO:356501945107:002a, OpenKIM-MO:356501945107:002, OpenKIM-MD:249792265679:002, OpenKIM-MO:616482358807:002a, OpenKIM-MO:616482358807:002, OpenKIM-MD:249792265679:002, OpenKIM-MO:101997554790:002a, OpenKIM-MO:101997554790:002, OpenKIM-MD:249792265679:002, OpenKIM-MO:307252285625:001a, OpenKIM-MO:307252285625:001, OpenKIM-MD:249792265679:002, OpenKIM-MO:878712978062:003a, OpenKIM-MO:878712978062:003, OpenKIM-MD:113599595631:003, OpenKIM-MO:353393547686:002a, OpenKIM-MO:353393547686:002, OpenKIM-MD:249792265679:002, OpenKIM-MO:086900950763:002a, OpenKIM-MO:086900950763:002, OpenKIM-MD:249792265679:002, OpenKIM-MO:114797992931:000a, OpenKIM-MO:114797992931:000, OpenKIM-MD:120291908751:005, OpenKIM-MO:104806802344:005a, OpenKIM-MO:104806802344:005b, OpenKIM-MO:104806802344:005, OpenKIM-MD:120291908751:005, OpenKIM-MO:108983864770:005a, OpenKIM-MO:108983864770:005b, OpenKIM-MO:108983864770:005, OpenKIM-MD:120291908751:005}, 
Pt \cite{OpenKIM-MO:601539325066:000a, OpenKIM-MO:601539325066:000b, OpenKIM-MO:601539325066:000, OpenKIM-MD:120291908751:005, OpenKIM-MO:102190350384:005a, OpenKIM-MO:102190350384:005b, OpenKIM-MO:102190350384:005, OpenKIM-MD:120291908751:005, OpenKIM-MO:793141037706:002a, OpenKIM-MO:793141037706:002, OpenKIM-MD:249792265679:002, OpenKIM-MO:020840179467:002a, OpenKIM-MO:020840179467:002, OpenKIM-MD:249792265679:002, OpenKIM-MO:115316750986:001a, OpenKIM-MO:115316750986:001, OpenKIM-MD:128315414717:004, OpenKIM-MO:070797404269:002a, OpenKIM-MO:070797404269:002, OpenKIM-MD:249792265679:002, OpenKIM-MO:545073984441:002a, OpenKIM-MO:545073984441:002, OpenKIM-MD:249792265679:002, OpenKIM-MO:716623333967:002a, OpenKIM-MO:716623333967:002, OpenKIM-MD:249792265679:002, OpenKIM-MO:831380044253:002a, OpenKIM-MO:831380044253:002, OpenKIM-MD:249792265679:002, OpenKIM-MO:637493005914:001a, OpenKIM-MO:637493005914:001, OpenKIM-MD:128315414717:004, OpenKIM-MO:343168101490:002a, OpenKIM-MO:343168101490:002, OpenKIM-MD:249792265679:002, OpenKIM-MO:280985530673:002a, OpenKIM-MO:280985530673:002, OpenKIM-MD:249792265679:002, OpenKIM-MO:912978207512:002a, OpenKIM-MO:912978207512:002, OpenKIM-MD:249792265679:002, OpenKIM-MO:534993486058:001a, OpenKIM-MO:534993486058:001, OpenKIM-MD:249792265679:002, OpenKIM-MO:500121566391:004a, OpenKIM-MO:500121566391:004b, OpenKIM-MO:500121566391:004c, OpenKIM-MO:500121566391:004, OpenKIM-MD:077075034781:005}, 
Rh \cite{OpenKIM-SM:306597220004:000a, OpenKIM-SM:306597220004:000, OpenKIM-SM:066295357485:000a, OpenKIM-SM:066295357485:000}

\begin{singlespace}
	\printbibliography[heading=subbibintoc,
					   title=Supplementary Information References]
\end{singlespace}

\end{refcontext}

\end{document}